\documentstyle[prd,aps]{revtex}
\begin{document}
\draft
\tighten
\title{     The Post Minkowskii Expansion of General Relativity }
\author{    Steven Detweiler  and Lee H. Brown Jr. }
\address{   Department of Physics, University of Florida,
            Gainesville, FL 32605 }
\date{July 15, 1997}
\maketitle
\begin{abstract}
A post-Minkowskii approximation of general relativity is described as
a power series expansion in $G$, Newton's gravitational constant.
Material sources are hidden behind boundaries, and only the vacuum
Einstein equations are considered.  An iterative procedure is
outlined which, in one complete step, takes any approximate solution
of the Einstein equations and produces a new approximation which has
the error decreased by a factor of $G$.  Each step in the procedure
consists of three parts: first the equations of motion are used to
update the trajectories of the boundaries; then the field equations
are solved using a retarded Green function for Minkowskii space;
finally a gauge transformation is performed which makes the geometry
well behaved at future null infinity. Differences between this
approach to the Einstein equations and similar ones are that we use a
general (non-harmonic) gauge and formulate the procedure in a
constructive manner which emphasizes its suitability for
implementation on a computer.
\end{abstract}
%\narrowtext
\section{Introduction}
Close binary systems are a possible source for the LIGO or VIRGO
gravitational wave detectors, but the non-linearities of the
Einstein equations have made such systems difficult subjects of
analyses.  Although, most notably, the post-Newtonian expansion
{\cite{Blanchet95,BlanchetDI95,BlanchetDIWW95,BlanchetIWW96,Will96}}
has now been used to calculate the gravitational wave form
resulting from the inspiralling evolution of the binary through
terms of order $(v/c)^5${\cite{Blanchet96a}}.

In anticipation of the day when such analytical results can be
pushed no further, we have developed a purely constructive,
iterative approach that allows for straightforward numerical
implementation.  We present a variation of a post-Minkowskii
expansion for relativistic systems.  Formulated as an iterative
procedure, it takes an approximate solution to the Einstein
equations and produces a better one---the error decreases by a
factor of $G$ with every iteration. The procedure reproduces the
standard lowest order post-Newtonian equations of motion and the
traditional quadrupole formula for gravitational waves.  But
higher orders are generally too complicated for analytical work
without the slow-motion assumption of the post-Newtonian
approximation.  However, on a computer the $n$th iteration of the
process is no more difficult than the first, and it is there
where we envisage putting this procedure to good use.

Two sources in a close binary system are likely to contain
regions of extreme curvature. However, when the sources are far
enough apart the multipole moments of the individual sources
depend very little on the presence of a companion and the
evolution, say, of two neutron stars differs only slightly from
that of two black holes or any other small, massive objects. In
our approach a boundary surrounds each source---and we focus on
the vacuum Einstein equations in the region outside of these
boundaries.

In many ways our approach is a combination of methods developed
by others.  It starts with a formal expansion of the Einstein
equations in powers of Newton's gravitational constant, $G$, in a
manner similar to Kerr{\cite{Kerr59}}.  The first order results
are formulated using a multipole expansion in terms of symmetric
trace free tensors described by Pirani{\cite{Pirani}},
Thorne{\cite{ThorneRMP}} and Blanchet and
Damour{\cite{BlanchetDphil}}. Particularly at first order our
results formally appear quite similar to
Thorne's{\cite{ThorneRMP}} analysis of linearized gravity, but our
multipole expansion is done about the moving boundaries which
hide the sources. At higher orders, we mimic the approach of
Blanchet and Damour{\cite{BlanchetDphil,BlanchetD89}} but allow
for a general (not harmonic) gauge which is restricted only
enough to be well behaved at future null infinity. Throughout our
analysis, we rely heavily upon the rigorous mathematical results
provided by Blanchet and
Damour{\cite{BlanchetDphil,BlanchetD89,Blanchet87}}.

One original aspect of our approach is the freedom from gauge
requirements---at least within the constraints of the metric
being considered a tensor field on a flat background space-time.
Also original is our analysis of the equations of motion which,
at lowest order, is similar to that of Bel {\it et
al.}{\cite{Bel}} except that we enclose the sources within
boundaries and avoid any formally divergent integrals.  At each
order of the approximation we find the equations of motion as
consequences of the desire to have the world line of the center
of the boundaries actually match the trajectory of the physical
source which is contained inside.

Also, we often use flat-space outgoing-null spherical coordinates
attached to the accelerating world lines.  These provide globally
well behaved coordinate systems and seem particularly well suited
for problems involving black holes as long as the relevant
accelerations are small, $m\dot v \ll 1$.

In \S{\ref{basic}} a summary of our assumptions and approximations
is presented and followed by descriptions of the notation,
coordinate systems and formulation of the general Einstein
equations which we use. \S{\ref{iter}} gives a description of the
iterative procedure along with specific details of how to start
the process at the first order, and how to iterate the field
equations at $n$th order.  But, for the field equations to have a
well behaved solution it is first necessary to iterate the
equations of motion as described in
\S{\ref{dynamics}}.  The behavior of the gravitational field at
future-null infinity is discussed in
\S{\ref{infinity}}.  Some details are relegated to the Appendix
including a description of the retarded Poincare
transformation---a convenient generalization of the Lorentz
transformation which relates outgoing-null spherical coordinates
centered on different world lines.

Our process is described in a manner that should make
implementation on a computer straightforward, particularly for
the analysis of the binary inspiral problem from the time of the
post-Newtonian applicability perhaps down to the innermost stable
circular orbit, where speeds could be $\sim c/2$. We expect this
approach to fail when the evolutionary time scale is comparable to
the dynamical time scale.

\section{Basic formulation} \label{background}
{\label{basic}}
\subsection{Assumptions and approximations}
We assume that Newton's gravitational constant, $G$, is small,
but not necessarily very small, when the units are such that a
characteristic mass and distance are of order 1 for a specific
problem.  This is essentially equivalent to assuming that the
metric of space-time deviates only modestly from being flat. We
also use units where the speed of light is unity---the
characteristic time corresponds to the light travel time across
the characteristic distance. We specifically do not assume that
the speeds are small, but we do assume that accelerations are,
$\dot{v}^a = O(G)$.  This is consistent with the assumptions of
weak fields---when fast, not-strong-field sources interact their
accelerations are small because the fields are weak.

We deal only with the vacuum Einstein equations and assume that
any material sources are shrouded by boundaries at surfaces of
constant $r$ in outgoing-null spherical coordinates centered on
each source.  Some geometrical data can be given on the boundary
in order to distinguish, say, a system involving black holes from
one of neutron stars. This distinction may be difficult to
implement on a computer, and numerical results may only be of
general validity and unable to carefully examine features of the
geometry which depend upon the detailed nature of the sources.
But, at least in principle, if this process converges to an
accurate solution of the Einstein equations, then the detailed
nature of the true physical geometry at the boundary can be
examined to see whether it is physically consistent with any
particular source of interest.

We have only modest gauge restrictions which are imposed to keep
the geometry well behaved at future null infinity. In particular,
we do not require the harmonic gauge.

In \S{\ref{dynamics}} we assume that the mass monopole moment,
${\cal A}$, is larger than all others by $O(G)$.  This is not a
required assumption, but it is reasonable on physical grounds and
allow us to describe the dynamical equations in familiar terms.

And in \S{\ref{infinity}} we assume that there exists an initial
spacelike hypersurface in the past of which the geometry is an
exact solution of the Einstein equations.  The geometrical data
on this hypersurface must satisfy minimal gauge requirements at
large values of $r$.  And the first order approximation to the
Einstein equations, $h_1^{ab}$, must match up smoothly with the
data on this initial hypersurface.

\subsection{Notation} {\label{notation}}

This formalism considers a gravitational field as a tensor field
on a flat Minkowskii background.  The mathematical notation is
that of flat-space tensor analysis with Minkowskii coordinates
$(t,x,y,z)$, along with the usual flat metric, $\eta_{ab}$, and
its inverse, $\eta^{ab}$, which are both $(-1,1,1,1)$ down the
diagonal and zero elsewhere. The operator $\nabla_a$ is just the
usual derivative operator of flat space, and $\nabla^2 \equiv
\nabla_a \nabla^a$.  Only $\eta_{ab}$ and $\eta^{ab}$ are used to
lower and raise tensor indices, and a tensor index {\em always}
refers to a component in Minkowskii coordinates.

For a binary system the positions of the two sources are
described by two different world lines of the form $z^a(s)$
parameterized by the Minkowskii proper time, $s$. In the
discussion the focus is usually upon only one world line at a
time with four-velocity, $v^a$.  It is convenient to use $z_s^a
\equiv z^a(s)$.

Occasionally we use outgoing-null spherical coordinates
$(s,r,\theta,\phi)$ based on the world line $z^a(s)$. The
scalar field $s$ is defined at an event, ${\cal P}$, of
Minkowskii space by $s({\cal P}) = s({\cal Q})$, where ${\cal Q}$ is
the vertex of the future null cone from $z^a(s)$ which
contains ${\cal P}$; it is convenient to use $s_x \equiv
s(x^a)$.  In a similar manner $v^a$, as well as any other
tensor defined along the world line, can be promoted to a
tensor field over all space-time by parallel transport of
$v^a({\cal Q})$ up the future null cone of ${\cal Q}$ to ${\cal P}$.

Any field derived from a tensor defined along the world
line, has a particularly simple expression for its derivative.
For example
\begin{equation}
  \nabla_b v^a = - k_b \dot v^a
\end{equation}
where
\begin{equation}
  k_a \equiv - \nabla_a s,
\end{equation}
and a dot denotes a derivative with respect to the retarded time
$s$, so that
\begin{equation}
  \dot v^a = v^b \nabla_b v^a = dv^a/ds.
\end{equation}

The quantity $r$, in the outgoing-null coordinates, is the
Minkowskii spatial distance between ${\cal Q}$ and ${\cal P}$ as
measured in the instantaneous rest frame of the world line at
${\cal Q}$,
\begin{equation}
  r(x^a) = - v_a [x^a - z^a({\cal Q})].
{\label{rdef}}
\end{equation}
Pirani{\cite{Pirani}} shows that $k^a$ is the null vector
field
\begin{equation}
  k^a = (x^a - z^a({\cal Q})) / r,
\end{equation}
pointing from ${\cal Q}$ to ${\cal P}$.  Also
\begin{equation}
  k^a v_a = -1,
\end{equation}
and
\begin{eqnarray}
  \nabla_b r & = &- v_b + k_b (1 + r k_a \dot{v}^a)
\nonumber \\
  & = & n_b+ r k_b k_a \dot{v}^a
{\label{GFB-4}}
\end{eqnarray}
where $n_b \equiv k_b - v_b$ is an outward-pointing, spatial unit
vector.  It is useful to know that
\begin{equation}
  r \nabla_a k_b = \eta_{ab}  +  v_a k_b  +  v_b k_a
             - k_a k_b (1 +  r k_c \dot v^c).
{\label{gradk}}
\end{equation}

The angles $\theta$ and $\phi$ at ${\cal P}$ are defined in the
usual way with the origin at ${\cal Q}$ and with a set of
orthonormal basis vectors which is Fermi-Walker transported along
the world line and parallel transported up the null cone.

The projection operator onto the spatial three manifold
instantaneously orthogonal to $v^a$ at ${\cal Q}$ is $f^{ab} \equiv
\eta^{ab} + v^a v^b$, and the alternating tensor orthogonal to
$v^a$ is $\epsilon^{abc} \equiv \epsilon^{abcd} v_d$.  But, note
that when $f^{ab}$ is promoted to a tensor field via parallel
transport up the future null cone it does not become the spatial
three metric of a constant $t$ surface if the world line is
accelerating.

The description of tensor multipole moments often requires a
large set of, say, $l$ indices; we follow Blanchet and
Damour{\cite{BlanchetDphil}} and define a tensor multi-index,
${}_{L} \equiv {}_{d_1 \ldots d_l}$ to denote a succession of $l$
space-time indices.  We use $N^{L} \equiv n^{d1} \ldots n^{d_l}$
for the tensor outer product of $l$ vectors, $n^d$, and
$\nabla_{L} \equiv \nabla_{d_1} \ldots \nabla_{d_l}$ for a
succession of $l$ derivatives.  Often we sum over $l$ from $0$ to
$\infty$, and this summation is assumed to converge without
justification being given. Sometimes a set of tensor indices are
symmetric, spatial with respect to $v^a(s)$ and completely trace
free, these are referred to as being SSTF.  If a tensor has all
of its indices SSTF and is a function of only $s$, then it is
denoted by a capital, script base letter.  Damour and
Iyer{\cite{DamourI91b}} give a host of useful formulae for
decomposing a tensor into SSTF parts.  We follow their notation
and equivalently denote the SSTF part of a tensor $A^L$ by ${\hat
A}^L \equiv A^{<L>} \equiv A^{<a_1\ldots a_l>}$.  Also, $[l/2]$
is just the largest integer less than or equal to $l/2$.

\subsection{The Einstein equations on a flat background manifold.}
{\label{pmexpan}}
A metric, $g_{ab}$, on a four dimensional space-time may be
considered as a two indexed, symmetric invertible tensor field on
a flat, background Minkowskii space.  It is convenient to define
$h^{ab}$ by
\begin{equation}
  \sqrt{-g}g^{ab} \equiv \eta^{ab}-h^{ab}
\end{equation}
and an Einstein tensor density as a functional of $h^{ab}$,
\begin{equation}
      E^{ab}(h) \equiv (-g)(2R^{ab}- g^{ab}R),
\end{equation}
so that the vacuum Einstein equation is
\begin{equation}
  E^{ab}(h) = 0.
\end{equation}
Landau and Lifshitz{\cite{LandL}} give an exact form for
$E^{ab}$; we write this as
\begin{eqnarray}
  E^{ab}(h) = -&\nabla^2  h^{ab}
     + \nabla^a\nabla_c h^{cb}
     + \nabla^b\nabla_c h^{ca}
\nonumber\\ &
  -\eta^{ab} \nabla_c\nabla_d h^{cd} - 16\pi\tau^{ab}(h)
{\label{defE}}
\end{eqnarray}
where
\begin{eqnarray}
  1&6&\pi \tau^{ab} (h)  \equiv  - 2\nabla_c h^{ab} \nabla_d h^{dc}
    - h^{ab} \nabla_c \nabla_d h^{cd}
\nonumber\\&&
    {}+ h^{ad} \nabla_d \nabla_c h^{cb}
    + h^{bd} \nabla_d \nabla_c h^{ca}
    + \nabla_d h^{ad} \nabla_c h^{bc}
\nonumber\\&&
    {}- h^{cd} \nabla_c \nabla_d h^{ab}
    + \nabla_c h^{ad} \nabla_d h^{bc}
    + 16\pi (-g) \tau^{ab}_{\rm LL}(h).
{\label{deftau}}
\end{eqnarray}
The quantity $\tau^{ab}_{\rm LL}(h)$ is the Landau-Lifshitz
pseudotensor, Eq.\ (96.9) in {\cite{LandL}} or Eq.\ (20.22)
in Misner {\it et al.}{\cite{MTW}}.

The Bianchi identity translated onto the flat background takes
the form
\begin{equation}
  \nabla_a E^{ab} =
        (\eta_c{}^b \Gamma^d_{da} - \Gamma^b_{ac}) E^{ac},
{\label{bianchi}}
\end{equation}
where $\Gamma^b_{ac}$ is the usual Christoffel symbol for the
real space-time metric, $g_{ab}$, and is $O(G)$.

\section{Iterative procedure} {\label{iter}}

We formally expand the gravitational field in powers of $G$:
$h_0^{ab}$ is zero, and at first order $h_1^{ab}= O(G^1)$ and
exactly matches the geometrical data on the initial hypersurface.
We iteratively assume that $E^{ab}(h_{n-1}) = O(G^n)$, with no
gauge restrictions on $h_{n-1}^{ab}$ (in particular it need not
be in the harmonic gauge), and look for a correction, $\delta
h_n^{ab} = O(G^n)$, such that
\begin{equation}
   h_n^{ab}(x) \equiv h_{n-1}^{ab}(x) + \delta h_n^{ab}(x;G)
{\label{expandh}}
\end{equation}
and $E^{ab}(h_n) = O(G^{n+1})$. The dependence of $\delta
h_n^{ab}$ on $G$ is allowed to be more complicated than just
being proportional to $G^n$, and in Eq.\ ({\ref{expandh}}) that
functional dependence is explicit---usually the dependence on $G$
is just understood.

For a given $h_{n-1}^{ab}$, the next order approximation follows
from a solution of
\begin{equation}
  \nabla^2 \delta h_n^{ab} = E^{ab}(h_{n-1}) + O(G^{n+1}),
  {\label{delhn}}
\end{equation}
for $\delta h_n^{ab} = O(G^n)$, with the additional restriction
that
\begin{equation}
  \nabla_a \delta h_n^{ab} = O (G^{n+1}).
  {\label{restrict}}
\end{equation}
That $h_{n-1}^{ab} + \delta h_n^{ab}$ is a more accurate solution
to the Einstein equations is revealed by substitution into
Eq.\ ({\ref{defE}}) resulting in
\begin{eqnarray}
    &E&^{ab}(h_{n-1} + \delta h_n)
\nonumber\\ && =
        [ E^{ab}(h_{n-1} + \delta h_n) - E^{ab}(h_{n-1}) ]
               + \nabla^2 \delta h_n^{ab} + O(G^{n+1})
\nonumber\\ && =
       \nabla^a\nabla_c \delta h_n^{cb}
     + \nabla^b\nabla_c \delta h_n^{ca}
     - \eta^{ab} \nabla_c \nabla_d \delta h_n^{cd}
\nonumber\\ && \quad
     - 16\pi [ \tau^{ab}(h_{n-1} + \delta h_n) -
           \tau^{ab}(h_{n-1}) ] + O(G^{n+1}).
{\label{Eabhnplus}}
\end{eqnarray}
The first equality follows from Eq.\ ({\ref{delhn}}), and the
second from Eq.\ ({\ref{defE}}).  With the restriction
({\ref{restrict}}) each of the first three terms on the right hand
side are $O(G^{n+1})$, and $\tau^{ab}(h)$ is quadratic in
$h^{ab}$ and its derivatives, so the fourth term is also
$O(G^{n+1})$. Thus
\begin{equation}
  E^{ab}(h_{n-1} + \delta h_n) = O (G^{n+1}),
\end{equation}
and one full step of the iteration consists of solving Eq.\
({\ref{delhn}}) with the restriction ({\ref{restrict}}).

The restriction ({\ref{restrict}}) should not be considered a
gauge condition.  After all, at the $n$th step $h_{n-1}^{ab}$
need not satisfy any particular gauge choice.
And the restriction does
not involve the residual value of $\nabla_a h_{n-1}^{ab}$ in any
manner.  Thus, at $n$th order there is no limitation upon
$\nabla_a h_{n}^{ab}$.  Also, the divergence of Eq.\
({\ref{delhn}}) with the Bianchi identity ({\ref{bianchi}})
implies that $\nabla^2 \nabla_a \delta h_n^{ab} = O(G^{n+1})$;
thus, with the proper choices of initial data and of boundary
conditions the restriction follows naturally from the wave
equation ({\ref{delhn}}).  To emphasize finally the gauge freedom
allowed, note that after each iteration is complete a
$\lambda_n^a = O(G^{n})$ gauge transformation can change the
metric by $h_n^{ab} \rightarrow h_n^{ab} + \partial
\lambda_n^{ab}$ where
\begin{equation}
  \partial \lambda_n^{ab} \equiv \nabla^a \lambda_n^b + \nabla^b
              \lambda_n^a - \eta^{ab} \nabla_c \lambda_n^c.
{\label{defpartial}}
\end{equation}
This transformation preserves the accuracy of the approximation
and changes $E^{ab}(h_n)$ only at $O(G^{n+1})$ and only through
the change in $\tau^{ab}(h_n)$.  Also at the $n\text{th}$ order,
the metric may be changed by the addition of a small, arbitrary
symmetric tensor, $\gamma_{n+1}^{ab} = O(G^{n+1})$; $h_n^{ab}
\rightarrow h_n^{ab} + \gamma_{n+1}^{ab}$ only changes
$E^{ab}(h_n)$ at $O(G^{n+1})$.  Such changes are used in
\S{\ref{infinity}} to insure proper asymptotic behavior.

As an iterative procedure this is slightly more general than the
post-Minkowskii expansion of Blanchet and
Damour{\cite{BlanchetDphil}}.  If any initial approximation to the
Einstein equations has $E^{ab}(h_1) = O(\epsilon)$ for some small
epsilon, then after one step $E^{ab}(h_2) = O(\epsilon G)$. And
if $h_n^{ab}$ represents an exact solution, independent of gauge,
then the procedure terminates.

The remainder of this paper focuses on a specific, constructive
method for performing one full step.

\subsection{First order approximation}{\label{firstorder}}
We formally require at first order that $h_1^{ab}$ must match the
geometrical data on the initial hypersurface.  However this
limitation is only used in \S{\ref{infinity}}; in addition, for
applications on a computer we are unlikely to use exact
initial data.  For these reasons in this section we just look for an
$h_1^{ab}$ which resembles two moving sources and has $E^{ab}(h_1) =
O(G^2)$ but does not necessarily match onto good initial data.

A general multipole source, $M_1^{abL}(s)$, confined to a world
line, $z^a(s)$, has an $h_1^{ab} = \delta h_1^{ab}$ which both
satisfies
\begin{equation}
  \nabla^2 h_1^{ab} = - 4\pi \sum_{l=0}^{\infty} \int M_1^{abL}(s)
         \nabla_{L} \delta^4(x-z_s) \,ds
{\label{h1wave}}
\end{equation}
and also is of the general form (see
Appendix {\ref{greensfunct}})
\begin{eqnarray}
  h_1^{ab} & = & \sum_{l=0}^{\infty}
                 \int M_1^{abL} \nabla_{L} G(x-z_s) \,ds
\nonumber\\ & = & \sum_{l=0}^{\infty}
   \nabla_{L}[r^{-1} M_1^{abL}(s)].
{\label{h1soln}}
\end{eqnarray}
We write $M_1^{abL}$ as a sum of terms involving completely SSTF
tensors, which are $O(G)$ and functions only of $s$, in a manner
which parallels Thorne's Eq.\ (8.4){\cite{ThorneRMP}} or Blanchet
and Damour's Eq.\ (2.25){\cite{BlanchetDphil}}.  The resulting
most general multipole source has
\widetext
\begin{eqnarray}
  M_1^{abL} & \equiv & v^a v^b {\cal A}_1^{L}
    + 2 v^{(a} {\cal B}_1^{b)L}
    + 2 v^{(a} \epsilon^{b)}{}_q{}^{(d_l} {\cal C}_1^{L-1)q}
    + 2 v^{(a} f^{b)(d_l} {\cal D}_1^{L-1)}
    + f^{ab} {\cal E}_1^{L} + {\cal F}_1^{abL}
  \nonumber\\&&
    +  2 \epsilon^{a)}{}_q{}^{(d_l} {\cal G}_1^{L-1)q(b}
    + 2 f^{a)(d_l} {\cal H}_1^{L-1)(b}
    + 2 \epsilon^{a)q(d_l} {\cal J}_{1q}{}^{L-2} f^{d_{l-1})(b}
    + f^{a(d_l} {\cal K}_1^{L-2} f^{d_{l-1}) b}.
{\label{defM}}
\end{eqnarray}
The reverse parentheses imply symmetrization on $a$ and $b$.
Note that of these SSTF tensors, ${\cal B}_1^{L}$, ${\cal C}_1^{L}$,
${\cal H}_1^{L}$ and ${\cal J}_1^{L}$ always have $l \ge 1$, while
${\cal F}_1^{L}$ and ${\cal G}_1^{L}$ have $l \ge 2$.

But the $h_1^{ab}$ from Eq.\ ({\ref{h1soln}}) can represent the
first term in the expansion of the gravitational field of the
multipolar source of Eq.\ ({\ref{defM}}) only if the restriction
({\ref{restrict}}) is also satisfied.  A lengthy analysis of the
consequences of restriction ({\ref{restrict}}) invokes a useful
identity described in Appendix {\ref{usefulidentity}}, Eq.\
({\ref{identity}}), liberally uses $f^{ac} = \eta^{ac} + v^a v^c$
and $\nabla^2G = - 4\pi\delta^4(x-z_s)$ and ultimately results in
\begin{eqnarray}
  \nabla_a h_1^{ab}&&(x) = \sum_{l=0}^{\infty}
   \int \ [ v^b \dot{\cal A}_1^{L} +
         \dot{\cal B}_1^{bL} + \epsilon^b{}_q{}^{d_l}
         \dot{\cal C}_1^{qL-1} + f^{bd_l} \dot{\cal D}_1^{L-1}
     + v^b {\cal B}_1^{L} + f^{bd_l} {\cal E}_1^{L-1} + {\cal F}_1^{bL} +
     \epsilon^b{}_q{}^{d_l} {\cal G}_1^{qL-1}
\nonumber\\ &&
     + f^{bd_l} {\cal H}_1^{L-1} + v^b \ddot{\cal D}_1^{L} +
     \ddot{\cal H}_1^{bL} +
     \epsilon^{b}{}_q{}^{d_l}\ddot{\cal J}_1^{qL-1} +
     f^{bd_l}\ddot{\cal K}_1^{L-1} ] \nabla_{L} G(x-z_s) \, ds +
     O(G^2).
{\label{divh1}}
\end{eqnarray}  

%\narrowtext
\noindent
The completeness of
the decomposition of SSTF vectors and tensors allows the
decomposed parts in Eq.\ ({\ref{divh1}}) to be matched up
according to the location of the index ${}^b${}---whether it sits
on $v^b$, $f^{bd_l}$, $\epsilon^b{}_q{}^{d_l}$ or on an SSTF
object.  Thus the restriction ({\ref{restrict}}) requires that the
multipole source, $M_1^{abL}$, satisfy
\begin{eqnarray}
  \dot {\cal A}_1^L + {\cal B}_1^L + \ddot {\cal D}_1^L
& = & O(G^2)
\text{ [$v^b$], } {\label{adot}} \\
  \dot {\cal B}_1^L + {\cal F}_1^L + \ddot {\cal H}_1^L
& = & O(G^2)
\text{ [SSTF${}^b$], }{\label{bdot}} \\
  \dot {\cal C}_1^L + {\cal G}_1^L + \ddot {\cal J}_1^L
& = & O(G^2)
\text{ [${\epsilon^b}_q{}^{d_l}$], }{\label{cdot}} \\
  \dot {\cal D}_1^L + {\cal E}_1^L + {\cal H}_1^L
       + \ddot {\cal K}_1^L
& = & O(G^2)
\text{ [$f^{bd_l}$]. } {\label{enodot}}
\end{eqnarray}
These are closely related to Eqs.\ (8.5) of
Thorne{\cite{ThorneRMP}} and Eqs.\ (2.26)--(2.28) of Blanchet and
Damour{\cite{BlanchetDphil}}.

The multipole moments ${\cal A}_1^L \ldots {\cal K}_1^L$ represent the
distribution of a source as computed with respect to the world
line $z^a$.  And our ${\cal A}_1^L \ldots {\cal K}_1^L$ are closely
related to similar quantities used by Thorne{\cite{ThorneRMP}}.
The differences are that our ${\cal B}_1^L$, ${\cal C}_1^L$ and
${\cal D}_1^L$ correspond to the negative of his; our ${\cal G}_1^L$,
${\cal H}_1^L$ and ${\cal J}_1^L$ correspond to twice his; and our
multipoles are defined with respect to an arbitrary world line,
his are with respect to the origin of the Cartesian coordinates.
But this close correspondence easily allows us to follow his
analysis for physical interpretations of some of the lowest
multipole moments.  In particular $({\cal A}_1+\dot{\cal D}_1)v^a /4 +
({\cal B}_1^a+\dot{\cal H}_1^a)/4$ corresponds to the four momentum of
the source relative to the world line;
$v^{[a}({\cal A}_1^{b]}+\dot{\cal D}_1^{b]}-{\cal H}_1^{b]})/2 +
\epsilon^a{}_q{}^b({\cal C}_1^q +\dot {\cal J}_1^q)/2$ corresponds to
the total angular momentum of the source about the world line;
$-({\cal A}_1^a + \dot {\cal D}_1^a - {\cal H}_1^a)/({\cal A}_1 + \dot
{\cal D}_1)$ corresponds to the displacement of the center of mass,
away from the world line; and $({\cal C}_1^a + \dot{\cal J}_1^a)/2$
corresponds to the spin angular momentum.  Also $F_1^L$ and
$G_1^L$ each give a transverse (to $n^a$ and $v^a$ at large
distances) trace-free contribution to
$h_1^{ab}$ and, therefore, represent the moments of sources of
gravitational waves which are inside the boundaries.  These
interpretations reflect the algebraic resemblance of these
multipole moments, calculated via $h_1^{ab}$ on a boundary close
to a source as it moves along an accelerating world line, with
the corresponding physical quantities as usually defined at large
distances in linearized gravity.

Thorne{\cite{ThorneRMP}} describes a gauge transformation which
can be used to set all of ${\cal D}_1^L, {\cal H}_1^L,
{\cal J}_1^L$ and ${\cal K}_1^L$ to zero while still preserving
both Eqs.\ ({\ref{delhn}}) and ({\ref{restrict}}).  In this
gauge, then, any choice of the ${\cal A}_1^L,
{\cal B}_1^L, {\cal C}_1^L, {\cal E}_1^L, {\cal F}_1^L$ and
${\cal G}_1^L$ which satisfies Eqs.\
({\ref{adot}})--({\ref{enodot}}) could represent the first
order multipolar decomposition of a gravitational source.  A
simple, interesting source of this type has only ${\cal A}_1$
and ${\cal C}_1^a$ non-vanishing to give the source
both mass
and angular momentum but no additional structure.

\subsection{The $n$th order approximation}
{\label{deltah}}
In this section we discuss the procedure by which the next order
approximation is found, but we avoid issues of the behavior of
the approximation at large $r${}---that subject is analyzed in
\S{\ref{infinity}}.

A given $(n-1)$th order approximation to the Einstein equations
has $E_{n-1}^{ab} \equiv E^{ab}(h_{n-1}) = O(G^n)$.  A step in
the iterative procedure requires a solution of Eq.\
({\ref{delhn}}), with $\delta h_n^{ab} = O(G^n)$ which satisfies
the restriction ({\ref{restrict}}).  Following Blanchet and
Damour{\cite{BlanchetDphil,BlanchetD89}}, we represent
$\delta h_n^{ab}$ as the sum of a particular solution of the
inhomogeneous wave equation, $p_n^{ab}$, and a general solution
of the homogeneous wave equation, $q_n^{ab}$; thus,
\begin{equation}
  \delta h_n^{ab}  = p_n^{ab} + q_n^{ab},
\end{equation}
where
\begin{equation}
  \nabla^2 p_n^{ab}  =  E_{n-1}^{ab}
\end{equation}
and
\begin{equation}
  \nabla^2 q_n^{ab} = 0, \quad \text{except on the world line.}
{\label{delqn1}}
\end{equation}
And $q_n^{ab}$ is chosen so that
\begin{equation}
  \nabla_a p_n^{ab}+ \nabla_a q_n^{ab} = O(G^{n+1}).
{\label{divqn}}
\end{equation}

\subsubsection{The solution for $p_n^{ab}$  }
The quantity $p_n^{ab}$ is formally given, with a retarded Green
function, by
\begin{equation}
  p_n^{ab} \equiv - \frac{1}{4\pi}\int E_{n-1}^{ab}(x') G(x-x')
\,d^4x'.
{\label{pnab}}
\end{equation}
In a numerical implementation this evaluation is the single
computationally intensive element.  It would most likely be
performed through standard finite differencing of the
inhomogeneous wave equation.  The fields would propagate freely
through the interiors of the inner boundaries where the source,
$E_{n-1}^{ab}$, would be set to zero.  Boundary conditions for
$p_n^{ab}$ would only be imposed in the wave zone.  It is
convenient to interpret $p_n^{ab}$ as resulting from the
nonlinearity of the Einstein equations with $h_{n-1}^{ab}$
creating an effective stress-energy outside of the boundaries.
Similarly, $q_n^{ab}$ appears to come from an $n$th order
correction to the multipole moments of the source hidden behind
the boundary.  However, with a nonlinear theory, the split of
$\delta h_n^{ab}$ into two parts in a manner which depends upon
the location of the boundary is at least modestly arbitrary, and
these interpretations are only suggestive.

\subsubsection{The solution for $q_n^{ab}$}

As a solution of the homogeneous wave equation, except on the
world line, $q_n^{ab}$ has a general representation similar to
that of $h_1^{ab}$ given in \S{\ref{firstorder}},
\begin{equation}
  q_n^{ab}(x) = \sum_{l=0}^{\infty} \nabla_{L}[ r^{-1}
                 M_n^{abL}(s_x)].
{\label{qn}}
\end{equation}
The majority of this section shows how the SSTF components of
$M_n^{abL}$ are determined by the SSTF components of
$E_{n-1}^{ab}$ on the inner boundaries; the results are exhibited
in Eqs.\ ({\ref{Adot}})--({\ref{Enodot}}).

A consequence of the formal solution for $p_n^{ab}$ in terms of
the retarded Green function is that
\begin{eqnarray}
  \nabla_a p_n^{ab} = \frac{1}{4\pi}\int && \nabla_a'
   [E_{n-1}^{ab}(x') G(x-x') ] \,d^4x'
\nonumber\\ &&
   - \frac{1}{4\pi}\int \nabla_a'(E_{n-1}^{ab}) G(x-x') \,d^4x',
{\label{divpn}}
\end{eqnarray}
where use is made of integration by parts and the symmetry of the
retarded Green function.  From the Bianchi identity
({\ref{bianchi}}) and $\Gamma^a_{bc} = O(G)$, the second integrand
is $O(G^{n+1})$, and
\begin{eqnarray}
  \nabla_a p_n^{ab} =
     \frac{1}{4\pi}\int \nabla_a' [E^{ab}_{n-1}(x')
                    && G(x-x') ] \,d^4x'
\nonumber\\ &&
  {}+ O(G^{n+1}).
{\label{divpna}}
\end{eqnarray}
This four-volume integral reduces to boundary integrals about
each of the two sources and a third at large $r'$, which gives a
vanishing contribution, for $x$ fixed, as $r'$ goes to
infinity---this can be verified by a lengthy analysis which
starts with the multipolar expansion of the Green function
provided by Blanchet and Damour{\cite{BlanchetDphil}}.

From this last equation it is clear that outside of the inner
boundaries the $O(G^n)$ part of $\nabla_a p_n^{ab}$ is a
homogeneous solution of the vector wave equation.  And we must
obtain its multipolar decomposition in order to find the
corresponding $q_n^{ab}$, a homogeneous solution of the spin-two
wave equation which also happens to satisfy Eq.\ ({\ref{divqn}}).
The needed SSTF decomposition of $\nabla_a p_n^{ab}$ involves
surprising subtlety and concludes in Eqs.\ ({\ref{gradpn}}) and
({\ref{defK}}) below.

The boundary integral resulting from Eq.\ ({\ref{divpna}}) is
evaluated with $(s',r',\theta',\phi')$ coordinates and an
expansion in terms of SSTF tensors.  For simplicity we choose the
boundary surrounding each source to be a surface of constant
$r'=r_0$; this boundary is spherical in a comoving frame of
reference for non accelerating world lines and is appropriately
Lorentz contracted in a frame wherein the world line is moving.
Each boundary integral gives
\begin{eqnarray}
  \nabla_a p_n^{ab} =
    - \frac{1}{4\pi}\int \!\!
  \oint \nabla'_a r' \, E^{ab}_{n-1} && G(x-x')
       r_0'^2 \sin\theta' \, d\theta' \,d\phi' \, ds'
\nonumber\\ &&
  {}+ O(G^{n+1}).
{\label{GFE-4}}
\end{eqnarray}

A Taylor series expansion of the Green function about the world
line leads to the multipolar decomposition of the integrand in
Eq.\ ({\ref{GFE-4}}) on a two-sphere with constant retarded time,
$s'$.  But, the Green function is non-vanishing only on the past
null cone from the field point $x^a$; $\nabla_a p_n^{ab}$ picks
up a contribution only where the past null cone intersects the
three dimensional boundary, and this occurs for differing values
of $s$.  The point of intersection closest to the field point
will have null-coordinate value $s$, but the point of
intersection on the far side of the boundary will have
null-coordinate value approximately $s-2r_0$.  Thus, the best
two-sphere to use for the multipolar decomposition of $\nabla_a
p_n^{ab}$ is the one at $s' = s-r_0$.  And the Taylor series
expansion for the Green function about the point $z^a(s=s'+r')$ on
a hypersurface which is orthogonal to the world line has $x'^a =
z^a + r'n'^a + O(r'\dot{v}^a)$ with the right hand side evaluated
at $s'+r'$; thus,
\begin{eqnarray}
  G(x^a - x'^a)  = \sum_{l=0}^{\infty}
         \frac{(-r')^l}{l!}  N'^{L} \nabla_{L}
          && G(x^a-z^a(s'+r'))
\nonumber\\ &&
  {}+ O(G).
\end{eqnarray}
 Now, with
$\nabla'_a r'|_{s-r_0} = \nabla'_a r'|_{s} + O(G)$ on the
boundary and $s'\rightarrow s'-r_0$,
\begin{eqnarray}
  \nabla&&{}_a  p_n^{ab} = - \sum_{l=0}^{\infty} \int
           [ \frac{(-r_0)^{l+2}}{4\pi\,l!}
     \oint \nabla'_a r' \, E^{ab}_{n-1}(s'-r_0)
         N'^{L}
\nonumber\\ && \times \nabla_{L} G(x-z(s'))
         \sin\theta' \, d\theta' \,d\phi' ] \, ds' +
         O(G^{n+1}).
{\label{divpdelG}}
\end{eqnarray}

Below it is necessary that this integrand be in terms of SSTF
tensors.  To this end, $N^L$ is equal to a sum
of terms each of which is a symmetrized outer product of
projection operators $f^{a_1a_2}$ and of the SSTF
combinations ${\hat N}^L$; thus,
\begin{equation}
  N^L = \sum_{m=0}^{[l/2]} b_{l,m}
          f^{(a_1 a_2}  \ldots f^{a_{2m-1}a_{2m}}
          n^{<a_{2m+1}} \ldots n^{a_l>)},
\end{equation}
for a set of coefficients $b_{l,m}$ which are obtained in
Appendix~{\ref{blms}} and given in Eq.\ ({\ref{blm}}).
Now the substitution
$f^{a_1 a_2} = \eta^{a_1 a_2} + v^{a_1} v^{a_2}$ and the use of
Eq.\ ({\ref{defGreen}}) allows part of
the integrand of Eq.\ ({\ref{divpdelG}}) to be rewritten for all
$x \neq z_s$ as
\begin{eqnarray}
   N'^L \nabla_L G(x-z_s) & = &
          \sum_{m=0}^{[l/2]} b_{l,m}
          v^{a_1} \ldots v^{a_{2m}}
          n'^{<a_{2m+1}} \ldots n'^{a_{l}>}
\nonumber\\ && {} \times \nabla_L G(x-z_s).
\end{eqnarray}
Application of the useful identity ({\ref{identity}}) further
transforms the $v^{a_1} \ldots v^{a_{2m}}$ inside the integral
into proper-time derivatives, and rearrangement of
the summation ultimately yields
\begin{eqnarray}
  \nabla&&{}_a  p_n^{ab} = - \sum_{l,m=0}^\infty b_{l+2m, m}\int
         \frac{(-r_0)^{l+2m+2}}{4\pi(l+2m)!}
\nonumber\\&&
 \times   \frac{d^{2m}}{ds'^{2m}} [\oint \nabla'_a r' \,
         E^{ab}_{n-1}(s'-r_0) {\hat N}'^{L}
         \sin\theta' \, d\theta' \,d\phi' ]
\nonumber\\&&
 \times   \nabla_{L} G(x-z(s')) \, ds' +
         O(G^{n+1}),
\end{eqnarray}
which has the ${}^L$ indices SSTF.

Finally, the desired multipolar decomposition is
\begin{eqnarray}
  \nabla_a p_n^{ab} =
     - \sum_{l=0}^{\infty} \int
       K_n^{bL}(s)&& \nabla_{L} G(x-z_s) \, ds
\nonumber\\ && {}+ O(G^{n+1}),
{\label{gradpn}}
\end{eqnarray}
where we define
\begin{eqnarray}
  K_n^{bL}& (s) & {}\equiv  \sum_{m=0}^\infty b_{l+2m, m}
      \frac{(-r_0)^{l+2m+2}}{4\pi(l+2m)!}
\nonumber\\&&
          \times \frac{d^{2m}}{ds^{2m}}
          \oint_{s,r_0} \nabla_a r \,
           E_{n-1}^{ab}(s-r_0)
      {\hat N}^{L} \sin\theta \, d\theta \,d\phi.
{\label{defK}}
{\label{GFE-9}}
\end{eqnarray}
The ${}^L$ indices are explicitly SSTF, and
$K_n^{bL}$ has SSTF components ${\cal P}_n^{L}$,
${\cal Q}_n^{L}$, ${\cal R}_n^{L}$ and ${\cal S}_n^{L}$ defined from
\begin{eqnarray}
  K_n^{bL}(s) =
         v^b {\cal P}_n^{L}(s)
       &+& \epsilon^b{}_q{}^{(d_l} {\cal Q}_n^{L-1)q}(s)
\nonumber\\ &&
       {} + {\cal R}_n^{bL}(s)
       + f^{b<d_l} {\cal S}_n^{L-1>}(s).
{\label{defPQRS}}
\end{eqnarray}

With this decomposition of $\nabla_a p_n^{ab}$ in hand, we return
to the search for $q_n^{ab}$ of the form given in Eq.\ ({\ref{qn}})
which satisfies Eq.\ ({\ref{divqn}}).  This analysis closely
follows \S{\ref{firstorder}}.  Let
\begin{equation}
  q_n^{ab} = \sum_{l=0}^{\infty}
      \int M_n^{abL}(s) \nabla_{L} G(x-z_s) \, ds
{\label{qnab}}
\end{equation}
where the definition of $M_n^{abL}(s)$ in terms of ${\cal A}_n^{L}
\ldots {\cal K}_n^{L}$ is similar to Eq.\ ({\ref{defM}}).  Now,
when $\nabla_a q_n^{ab}$, as in Eq.\ ({\ref{divh1}}), is added to
$\nabla_a p_n^{ab}$, as in Eqs.\ ({\ref{gradpn}}) and
({\ref{defPQRS}}), and like terms are matched up, the result of
Eq.\ ({\ref{divqn}}) is
\begin{eqnarray}
  \dot{{\cal A}}_n^{L} + {\cal B}_n^{L} + \ddot{{\cal D}}_n^{L}
  - {\cal P}_n^{L}
& = & O(G^{n+1}), {\label{Adot}} \\
  \dot{{\cal B}}_n^{L} + {\cal F}_n^{L} + \ddot{{\cal H}}_n^{L}
  - {\cal R}_n^{L}
& = & O(G^{n+1}), {\label{Bdot}} \\
  \dot{{\cal C}}_n^{L} + {\cal G}_n^{L} + \ddot{{\cal J}}_n^{L}
  - {\cal Q}_n^{L}
& = & O(G^{n+1}), {\label{Cdot}} \\
   \dot{{\cal D}}_n^{L} + {\cal E}_n^{L} + {\cal H}_n^{L}
       + \ddot{{\cal K}}_n^{L}
  - {\cal S}_n^{L}
& = & O(G^{n+1}). {\label{Enodot}}
\end{eqnarray}
These form a set of coupled, linear, ordinary, inhomogeneous
differential equations for ${\cal A}_n^{L} \ldots {\cal K}_n^{L}$ with
sources involving ${\cal P}_n^{L} \ldots {\cal S}_n^{L}$.  And any
solution to these equations gives a corresponding $q_n^{ab}$ via
Eq.\ ({\ref{qn}}) which, along with $p_n^{ab}$ determines $\delta
h_n^{ab}$ and formally yields an improved, approximate solution
to the Einstein equations.

A particular solution to most of these equations results from
setting ${\cal A}_n^{L}$, ${\cal C}_n^{L}$, ${\cal D}_n^{L}$,
${\cal H}_n^{L}$, ${\cal J}_n^{L}$ and ${\cal K}_n^{L}$ to zero, then
\begin{eqnarray}
  {\cal B}_n^{L} &=& {\cal P}_n^{L}, \quad
 \quad  \quad {\cal F}_n^{L} = {\cal R}_n^{L} -  \dot{{\cal B}}_n^{L},
\nonumber\\
 {\cal G}_n^{L} & = &{\cal Q}_n^{L},
 \quad \text{and}
 \quad  {\cal E}_n^{L} =  {\cal S}_n^{L}.
{\label{EDl}}
\end{eqnarray}
The general solution to Eqs.\ ({\ref{Adot}})--({\ref{Enodot}}) is
this particular solution plus any homogeneous solution for the
${\cal A}_n^L \ldots {\cal K}_n^L$.  And a homogeneous solution added
in at the $n$th iteration is no different from starting the
entire iterative process with a slightly different choice for the
first order ${\cal A}_1^L \ldots {\cal K}_1^L$.

The specific solution of Eqs.\ ({\ref{Adot}})--({\ref{Enodot}}) to
be used should be determined by the physics of the interior.  For
example, the particular solution in Eq.\ ({\ref{EDl}}) is easy to
implement and leaves unchanged all of the mass and current
moments, ${\cal A}^L$ and ${\cal C}^L$ respectively.  This may be
loosely interpreted as the appropriate solution for a steady
object and is not unreasonable as a choice for any
astrophysically interesting source in a binary as long as tidal
effects are unimportant.  Also, this particular solution coupled
with the simple choice for the first order moments of only the
mass monopole, ${\cal A}_1$, and current dipole, ${\cal C}_1^b$, being
non-zero ought to be able to reproduce numerically the results of
the higher order post-Newtonian analyses which are currently
published and also contain no tidal effects.  But, to find the
solution appropriate for a tidally distorted star is more
difficult---$p_n^{ab}$ within the boundary creates tidal forces,
distorts the star and changes all of the moments,
${\cal A}^L \ldots {\cal K}^L$, in a manner which would need to be
determined. Or, for a black hole within the boundary a
perturbative analysis might be used to determine the appropriate
solution of Eq.\ ({\ref{Adot}})--({\ref{Enodot}}).  In any event, a
variety of different possibilities could be implemented; the
actual choice made should be specific to the physics of the
interior sources.

But, the particular solution, above, fails for the
low multipoles ${\cal B}_n$,
${\cal F}_n^{b}$ and ${\cal G}_n^{b}$ because these SSTF
tensors don't exist and can't satisfy
Eq.\ ({\ref{EDl}}).  Thus, we still must contend with
\begin{eqnarray}
  \dot{{\cal A}}_n + \ddot {\cal D}_n - {\cal P}_n
& = & O(G^{n+1}), {\label{mass}} \\
  \dot{{\cal A}}_n^b + {\cal B}_n^b + \ddot {\cal D}_n^b - {\cal P}_n^b
& = & O(G^{n+1}), {\label{momentum}} \\
  \dot{{\cal B}}_n^b + \ddot {\cal H}_n^b - {\cal R}_n^b
& = & O(G^{n+1}), {\label{force}} \\
  \dot{{\cal C}}_n^b + \ddot{{\cal J}}_n^b
  - {\cal Q}_n^b
& = & O(G^{n+1}). {\label{torque}}
\end{eqnarray}

These remaining ten ordinary linear differential equations have
simple interpretations.  $\dot {\cal A}_n + \ddot {\cal D}_n$ gives the
rate of change of the mass monopole moment, and ${\cal P}_n$ is
analogous to the rate energy flows into the source through the
boundary.  $\dot {\cal C}_n^b + \ddot {\cal J}_n^b$ is similar to the
rate of change of spin angular momentum, and ${\cal Q}_n^b$ is
analogous to the torque.  $\dot {\cal B}_n^b + \ddot {\cal H}_n^b$
gives the rate of change of momentum of the source with respect
to the world line, and ${\cal R}_n^b$ is analogous to the force.
$\dot {\cal A}_n^b + \ddot {\cal D}_n^b$ is closely related to the rate
of change of the dipole moment caused by the momentum of the
source with respect to the world line and by ${\cal P}_n^b$, which
has no common Newtonian analog.

Eqs.\ ({\ref{mass}}) and ({\ref{torque}}) may be integrated as
ordinary differential equations.  Then, after a proper time
$s = O(G^{-1})$, ${\cal A}_n$, for example,
will typically have grown large enough that the order of the
approximation will have decreased by one.  This is not
particularly troublesome, and just implies that to obtain an
approximate solution to the Einstein equations with
$E^{ab}(h_n) = O(G^N)$ after a time $s = O(G^{-m})$
requires that $n = N + m - 1$.  And for a binary system the
approximation loses one order only when the fractional change in
the mass of one of the components is $O(G^n)$.

But the physical interpretations of Eqs.\ ({\ref{momentum}}) and
({\ref{force}}) give cause for concern.  So far in this formalism
the world line has been given ahead of time.  And the changing
dipole moment and relative momentum of Eqs.\ ({\ref{momentum}})
and ({\ref{force}}) just reflect the fact that the true, physical
source is moving with respect to the predetermined world line.
But all of the moments of the source are calculated about the
world line which does not necessarily follow the center of mass
of the source.  And as the source drifts away from the world line
a rapidly growing number of multipole moments need to be
monitored to adequately describe the source. This would be a
disaster for any implementation.

Thus at the $n-1$ iteration, {\em before} getting to this stage, we
should have made certain that the trajectory of the
world line was chosen so that
\begin{equation}
  {\cal R}_n^b - \dot{\cal P}_n^b = O(G^{n+1}).
{\label{integcond}}
\end{equation}
Then with ${\cal B}_n^b = {\cal P}_n^b + O(G^{n+1})$, Eqs.\
({\ref{momentum}}) and ({\ref{force}}) are solved with all of
${\cal A}_n^b$, ${\cal D}_n^b$ and ${\cal H}_n^b$ being zero.  And with no
growing dipole moment the plethora of required moments is
avoided.

In the next section we show that Eq.\ ({\ref{integcond}}) is
essentially the $(n-1)$th order equation of motion of the source
and generally necessitates an $O(G^{n-1})$ adjustment of the
world line.

At this point (if not previously) one might wonder what the
effects of making a different choice for the radius of the
boundary, $r_0$, might be.  For example, if $r_0$ were increased
then Eq.\ ({\ref{pnab}}) shows that $p_n^{ab}$ would change by the
addition of a homogeneous solution to the wave equation whose
divergence would account for any consequent change in the
${\cal P}_n^{L} \ldots {\cal S}_n^{L}$.  Thus the change could be
absorbed by $q_n^{ab}$.

\section{Equations of motion}
{\label{dynamics}}

After the $n$th iteration of the field equations as described in
\S{\ref{deltah}}, both $h_n^{ab}$ and $z_{n-1}^a(s)$ are known.
Before iterating the field equations again, it is first
necessary to adjust the world line, $z_{n-1}^a(s) \rightarrow
z_n^a(s)$, in order to enforce the $n$th order equation of
motion, ${\cal R}_{n+1}^b - \dot{\cal P}_{n+1}^b = O(G^{n+1})$.

Care must be taken to insure that this adjustment is accomplished
while maintaining the order of accuracy of the current
approximation to the metric.  Thus, we require a satisfactory
method for pulling the self-field of a source along a new world
line.  An aid in this task is the invariance of the Einstein
equations under a Lorentz transformation, as formulated in
\S{\ref{pmexpan}}.  As we see below, the appropriate adjustment of
a world line necessarily involves changing its acceleration but
only by a small amount of $O(G^{n})$.  And, pulling the
self-field along a new world line is nearly, but not quite,
accomplished by a Lorentz transformation.  However, the retarded
Poincare transformation, described in Appendix {\ref{rpt}} as a
generalization of the Lorentz transformation, allows for a
time-dependent boost and is still adequately behaved
globally.  The Einstein equations are not strictly invariant
under a retarded Poincare transformation; but, as demonstrated
for the scalar wave equation in Appendix {\ref{rptscalar}}, they
are approximately so.  And the retarded Poincare transformation
is sufficient for the task.

First in this section, we show how to implement the retarded
Poincare transformation to pull the self-field of a source along
a new world line in a manner that maintains the accuracy of the
approximation to the Einstein equations.  Then, we show just how
the new world line is chosen to satisfy the $n$th order equation
of motion.

\subsection{Adjusting $h_n^{ab}$}
{\label{adjustingh}}
A retarded Poincare transformation, described in Appendix
{\ref{rpt}}, adjusts a world line by defining a new coordinate
system with
\begin{equation}
 y^{a'} = \Lambda^{a'}{}_b(s_x) x^b + \xi^{a'}(s_x),
{\label{adjust}}
\end{equation}
where $\Lambda^{a'}{}_b(s_x)$ is a matrix of the form of a
Lorentz transformation and a function of the retarded
time, $s_x$, at $x^b$; also $\xi^{a'}(s_x)$ satisfies Eq.\
({\ref{xidot}}).  The world line in these new coordinates is
\begin{equation}
 z_n^{a'} = \Lambda^{a'}{}_b(s_z) z_{n-1}^b + \xi^{a'}(s_z),
{\label{zan}}
\end{equation}
with the consequence, noted in Eq.\ ({\ref{vnew1}}), that
\begin{equation}
  v_{n}^{a'} = \Lambda^{a'}{}_b(s) v_{n-1}^b.
{\label{vnew}}
\end{equation}
We show below how to choose $\Lambda^{a'}{}_b(s)$ to determine a
new world line along which the equation of motion is
satisfied; this has $\Lambda^{a'}{}_b(s=0)=\delta^{a'}_b$,
to match smoothly to the initial data, and $\dot\Lambda =
O(G^n)$ to keep the adjustment small.

The field $h_n^{ab}$ is separated into a self field,
$h_A^{ab}$, and a background field, $h_B^{ab}$,
\begin{equation}
  h_n^{ab} = h_A^{ab} + h_B^{ab}
\end{equation}
where $h_A^{ab}$ contains at least the $O(G^1)$ part of the
source whose world line is under consideration, and $h_B^{ab}$
contains at least the $O(G^1)$ part of all other sources.  The
distribution of the remaining $O(G^m; m \ge 2)$ parts of
$h_n^{ab}$ between $h_A^{ab}$ and $h_B^{ab}$ is immaterial.

There is no unique way to pull the self-field $h_A^{ab}$ along
with the new world line.  But if $\Lambda^{a'}{}_b$ were constant
then the usual Lorentz boost would be required.  Thus, a natural
choice for the self-field associated with a world line adjusted
via a retarded Poincare transformation is
\begin{equation}
  h_{A\text{new}}^{a'b'}(y)  =
              \Lambda^{a'}{}_c \Lambda^{b'}{}_d h_A^{cd}(x),
{\label{hAnew}}
\end{equation}
and this seems particularly reasonable when
$\dot \Lambda^{b'}{}_c = O(G^{n})$ and is small.
Thus the new field, $h_{n\text{ew}}^{a'b'}(y)$, in the new
coordinates is chosen to be
\begin{eqnarray}
  h&&_{n\text{ew}}^{a'b'}(y) \equiv \Lambda^{a'}{}_c \Lambda^{b'}{}_d
          h_A^{cd}(x) + \delta^{a'}_c \delta^{b'}_d h_B^{cd}(y)
\nonumber\\ &&
  {} = \delta^{a'}_c \delta^{b'}_d h_n^{cd}(y)
          + [ \Lambda^{a'}{}_c \Lambda^{b'}{}_d h_A^{cd}(x)
          -   \delta^{a'}_c \delta^{b'}_d h_A^{cd}(y)]
{\label{defhnew}}
\end{eqnarray}
where $x$ is the function of $y$ consistent with the inverse of
Eq.\ ({\ref{adjust}}).  The background field $h_B^{ab}$ is the
same function of $y^{a'}$ as it was of $x^a$, while the self
field is simultaneously pulled along and boosted by the time
dependent Lorentz transformation, $\Lambda^{a'}{}_c(s)$.  This
choice is consistent with the derivation of Eq.\ ({\ref{eom1}})
below.

But, we must show that the nonlinearity of $E^{ab}(h)$ combines
with $\dot\Lambda^{a'}{}_b$ to change $E_n^{ab}$ only at
$O(G^{n+1})$.  The part of $h_{n\text{ew}}^{a'b'}(y)$ in square
brackets in Eq.\ ({\ref{defhnew}}) vanishes when $s$ is zero
(because $\Lambda^{a'}{}_b(s=0) = \delta^{a'}_b$), is
proportional to $h_A^{ab}$ and is, therefore, small and
$\sim s \dot\Lambda h_A = O(sG^{n+1})$.  Now,
$E^{a'b'}(h_{n\text{ew}}^{a'b'}(y) )$ can be expanded about its
value at $\delta^{a'}_c \delta^{b'}_d h_n^{cd}(y)$ and broken up
into the parts
\begin{eqnarray}
  E^{a'b'}(&&h_{n\text{ew}}^{a'b'}(y)) =
      E^{a'b'}[ \delta^{a'}_c \delta^{b'}_d h_n^{cd}(y) ]
\nonumber\\ &&
    {} + E_{\text{linear}}^{a'b'} [
               \Lambda^{a'}{}_c \Lambda^{b'}{}_d h_A^{cd}(x) -
                  \delta^{a'}_c \delta^{b'}_d h_A^{cd}(y) ]
\nonumber\\ &&
    {} + E_{\tau}^{a'b'}
    + O(s^2 G^{2n+2}),
{\label{Ehnew}}
\end{eqnarray}
where $E_{\text{linear}}^{a'b'}$ denotes the linear part of the
operator $E^{a'b'}$, from Eq.\ ({\ref{defE}}); and
$E_{\tau}^{a'b'}$ is the part derived from $\tau^{ab}$ which is
still linear in
$\Lambda^{a'}{}_c \Lambda^{b'}{}_d h_A^{cd}(x)
  - \delta^{a'}_c \delta^{b'}_d h_A^{cd}(y)$
but also depends upon $\delta^{a'}_c \delta^{b'}_d h_n^{cd}(y)$.
The $O(s^2 G^{2n+2})$ terms remaining are at least quadratic in
$\Lambda^{a'}{}_c \Lambda^{b'}{}_d h_A^{cd}(x) - \delta^{a'}_c
\delta^{b'}_d h_A^{cd}(y)$.  To observe how well
$h_{n\text{ew}}^{a'b'}(y)$ satisfies the Einstein equations, we
analyze Eq.\ ({\ref{Ehnew}}) term by term.

The first term is $O(G^{n+1})$ by assumption.

The functional argument of the second term consists of the
difference of two parts, each of which is $O(G)$, but whose
difference is $O(sG^{n+1})$.  The $O(G)$ piece of each part is a
solution to the linear Einstein equations.  And the difference of
the $O(G^2)$ pieces of the two parts is only $O(sG^{n+2})$.
Thus this second term is $O(sG^{n+2})$.

The third term consists of a sum of terms each of which is of the
order of at least the product of
$\delta^{a'}_c \delta^{b'}_d h_n^{cd}(y)$
with $\Lambda^{a'}{}_c \Lambda^{b'}{}_d h_A^{cd}(x)
  - \delta^{a'}_c \delta^{b'}_d h_A^{cd}(y)$;
the former is $O(G)$, the latter is
$O(sG^{n+1})$.  Thus the third term is $O(sG^{n+2})$.

All together then
\begin{equation}
  E^{a'b'}[h_{n\text{ew}}^{a'b'}(y)] = O(G^{n+1}) + O(sG^{n+2}).
{\label{errorgrow2}}
\end{equation}
And we see that while $s = O(G^{-1})$ the error is $O(G^{n+1})$,
after that the order of the approximation decreases by one in a
manner similar to \S{\ref{deltah}}.  This is not a severe
limitation on applications of this method to binary systems,
where the radius of the boundary can be chosen to be of the same
order of magnitude as the separation between the components, $R$.
For a total mass of the system, $M$, and a typical speed, $V$,
$GM/R \approx (V/c)^2 = O(G)$.  The order of the approximation
decreases by one only when $sc/R = O(G^{-2})$, and this occurs
when $sV \approx R(c/V)^3$.  Thus the binary must orbit on the
order of $(c/V)^3$ times before the order of approximation
decreases.

\subsection{Adjusting the world line}
{\label{worldline}}

At this stage of an iterative step, we know $z_{n-1}^a(s)$,
$h_A^{ab}(x)$ and $h_B^{ab}(x)$.  This section shows how a
change in the acceleration of the world line, effected by a
retarded Poincare transformation with non-vanishing
$\dot\Lambda^{a'}{}_b(s)$, insures that the $n$th order equation
of motion is satisfied.

For simplicity we assume that out of the first order moments,
${\cal A}_1^{L} \ldots {\cal K}_1^{L}$, only ${\cal A}_1$ is $O(G)$, and
all of the others are smaller and $O(G^2)$.  This is a reasonable
physical assumption for astrophysical objects, and it should be
clear how to generalize this analysis if one desires to examine,
say, spin-orbit coupling by including ${\cal C}_1^b$ terms here, as
well.

When the acceleration of the world line is changed, the dominant
effect on ${\cal R}_n^b$ and ${\cal P}_n^b$ arises from their parts
which are linear in ${\cal A}_1$, depend upon $\dot{v}^a$, or
$\ddot{v}^a$, and are $O(G^2)$.  We call these parts
${\cal R}^b_{{\cal A}}$ and ${\cal P}^b_{{\cal A}}$; and the next paragraph
evaluates ${\cal R}^b_{{\cal A}} - \dot{\cal P}^b_{{\cal A}}$ with a
conclusion in Eq.\ ({\ref{ma}}).

Both ${\cal R}^b_{{\cal A}}$ and ${\cal P}^b_{{\cal A}}$ depend upon
$E^{ab}_{{\cal A}}$, the part of $E_{n-1}^{ab}$ which is linear in
${\cal A}_1$ and $O(G^2)$.  From Eqs.\ ({\ref{defE}}),
({\ref{h1wave}}) and ({\ref{h1soln}}) it follows that
\begin{eqnarray}
  E_{{\cal A}}^{ab} & = - 2&r^{-1}{\cal A}_1 [ \ddot{v}^{(a} k^{b)}
   +r^{-1} \dot v^{(a} (k^{b)} - v^{b)}) + \dot{v}^{(a} k^{b)}
   k_c \dot v^c
\nonumber\\ &&
{} - \case{1}{2}\eta^{ab} (k_c \ddot{v}^c + r^{-1} k_c \dot{v}^c
        + (k_c \dot{v}^c)^2) ].
{\label{GFJ-2b}}
\end{eqnarray}
The evaluation of ${\cal R}_{{\cal A}}^b$ and ${\cal P}_{{\cal A}}^b$ from
Eqs.\ ({\ref{defK}}) and ({\ref{defPQRS}}) requires the two
integrals
\begin{equation}
    \frac{1}{4\pi}\oint_{s,r_0}
    \nabla_a r E_{{\cal A}}^{ab} r_0^2
    \sin\theta \, d\theta \, d\phi
     =  - {\cal A}_1 (\dot v^b + r_0 \ddot v^b) + O(G^3)
{\label{GFJ-8}}
\end{equation}
and
\begin{equation}
    \frac{-1}{4\pi} \oint_{s,r_0} \nabla_a r \,
          E_{{\cal A}}^{ab} n^d r^3 \sin\theta \, d\theta \, d\phi
       =  \case{1}{3} {\cal A}_1 r_0^2 v^b \ddot v^d + O(G^3),
{\label{GFJ-8b}}
\end{equation}
where use is made of the fact that
$v_b \ddot v^b = - \dot v_b \dot v^b = O(G^2)$.
Now Eq.\ ({\ref{GFJ-8}}),
along with $b_{2m,m}=1/(2m+1)$ from Eq.\ ({\ref{b2mm}}), yields
\begin{eqnarray}
  {\cal R}_{{\cal A}}^b(s)
     &=& - {\cal A}_1 \sum_{m=0}^\infty
       \left[
         \frac{r_0^{2m}}{(2m+1)!}
         \frac{d^{2m}}{ds^{2m}}
         (\dot v^b + r_0 \ddot v^b)
       \right]_{s-r_0}
\nonumber\\ &&
       {}+ O(G^3).
{\label{ra2}}
\end{eqnarray}
That the right hand side here is evaluated at $s-r_0$ is a
complicating consequence of Eq.\ ({\ref{defK}}).  Also, from
Eq.\ ({\ref{GFJ-8b}}) along with $b_{1+2m,m}=3/(2m+3)$ from
Eq.\ ({\ref{b12mm}}), it follows that
\begin{eqnarray}
  {\cal P}_{{\cal A}}^b(s)
         &=& {\cal A}_1 \sum_{m=0}^\infty
       \left[
         \frac{r_0^{2m+2}}{(2m+3)(2m+1)!}
         \frac{d^{2m}}{ds^{2m}}
           \ddot v^b
       \right]_{s-r_0}
\nonumber\\ &&
       {} + O(G^3).
{\label{pa2}}
\end{eqnarray}
These two equations combine to yield
\widetext
\begin{eqnarray}
  ({\cal R}_{{\cal A}}^b - \dot {\cal P}_{{\cal A}}^b)_{s}
      = - ({\cal A}_1 \dot v^b )_{s-r_0}
    & - & {\cal A}_1 \sum_{m=0}^\infty
         \left[
               \frac{r_0^{2m+2}}{(2m+3)!}
               \frac{d^{2m}}{ds^{2m}}\overdots v^b
        \right.
\nonumber\\ &&
   \left.  {} +  \frac{r_0^{2m+1}}{(2m+1)!}
               \frac{d^{2m}}{ds^{2m}}\ddot v^b
      + \frac{r_0^{2m+2}}{(2m+3)(2m+1)!}
               \frac{d^{2m}}{ds^{2m}}\overdots v^b
      \right]_{s-r_0}
      + O(G^3);
\end{eqnarray}
%\narrowtext
\noindent
the first two terms inside the summation come from the $\dot v^b$
and $\ddot v^b$ parts of ${\cal R}_{{\cal A}}^b$, respectively; the
third term comes from $\dot {\cal P}_{{\cal A}}^b$.  The two
$\overdots v^b$ terms add directly, and the entire expression
simplifies remarkably to
\begin{equation}
  ({\cal R}_{{\cal A}}^b - \dot {\cal P}_{{\cal A}}^b)_{s}
     = - {\cal A}_1 \sum_{m=0}^\infty
           \left(
             \frac{r_0^m}{m!} \frac{d^{m}}{ds^{m}}\dot v^b
           \right)_{s-r_0}
           + O(G^3).
\end{equation}
Finally, the right hand side is a Taylor series expansion of
$-{\cal A}_1 \dot v^b$ about $s-r_0$ but evaluated at $s$ so that
\begin{equation}
  {\cal R}_{{\cal A}}^b - \dot {\cal P}_{{\cal A}}^b
           = - {\cal A}_1 \dot v^b + O(G^3),
{\label{ma}}
\end{equation}
and both sides of this equation are now evaluated at the same $s$.
The surprising simplicity of this last result might imply that a
substantially more straightforward derivation could be found.

A retarded Poincare transformation, with Eq.\ ({\ref{vnew}}),
effects the acceleration of the world line by
\begin{equation}
     \dot v_n^{b'} = \Lambda^{b'}{}_a \dot v^a_{n-1}
                   + \dot\Lambda^{b'}{}_a v^a_{n-1}.
{\label{dotv}}
\end{equation}
Now, we assume that $\Lambda^{b'}{}_a(s)$ has been determined
consistently with the equation of motion for all $s$
up to some value $s_0$.  Then both ${\cal R}_{n+1}^{b'}$ and
$\dot{\cal P}_{n+1}^{b'}$ at $s_0$ can be found from Eqs.\
({\ref{defK}}) and ({\ref{defPQRS}}) along with the temporary
assumption that $\dot\Lambda^{b'}{}_a(s_0) = 0$.  And we define
\begin{equation}
  F^{b'} \equiv ({\cal R}_{n+1}^{b'} - \dot{\cal P}_{n+1}^{b'})_{s_0}
       \quad \text{with} \quad 	\dot\Lambda^{b'}{}_a(s_0) = 0.
{\label{Fdef}}
\end{equation}
But, a value for
$\dot\Lambda^{b'}{}_a(s_0)$ of $O(G^n)$ changes
$({\cal R}_{n+1}^{b'}-\dot{\cal P}_{n+1}^{b'})_{s_0}$ by
$-{\cal A}_1\dot\Lambda^{b'}{}_a v_{n-1}^a+O(G^{n+2})$, from Eqs.\
({\ref{ma}}) and ({\ref{dotv}}). Thus if we
choose at $s_0$ that
\begin{equation}
  {\cal A}_1 \dot\Lambda^{b'}{}_a(s_0) v^a_{n-1} = F^{b'}
{\label{eom1}}
\end{equation}
then the adjusted world line will satisfy the $n$th order
equation of motion at $s_0$ as well.

This differential equation for $\Lambda^{b'}{}_a$ is consistent
with $\dot\Lambda^{b'}{}_a = O(G^{n})$, as promised in
\S{\ref{adjustingh}}, and gives three equations for
$\Lambda^{b'}{}_a$ (the ${}^{b'}$ index is orthogonal to
$v^{b'}$); the remainder of $\Lambda^{b'}{}_a$ is determined by
the requirement of Fermi-Walker transport Eq.\ ({\ref{fermiw2}}).

With Eq.\ ({\ref{dotv}}) the differential equation can be
rewritten as
\begin{equation}
  {\cal A}_1 \dot v_{n}^{b'} = {\cal A}_1 \Lambda^{b'}{}_a \dot v_{n-1}^a
                 + F^{b'},
{\label{eom3}}
\end{equation}
which has the expected form for an iteration of the equation of
motion with $F^{b'}$ a residual force remaining on world line
$z_{n-1}(s)$.  It is not difficult to show that at the first
order this equation of motion is equivalent to the usual
post-Newtonian result as presented by Bel {\it et
al.}{\cite{Bel}}.

\section{Behavior at future null infinity}{\label{infinity}}
Now we reconsider the iterative procedure outlined in
\S{\ref{iter}} with particular attention given to the limit of
large $r$ while $s=t-r$ is held constant.  Thus we consider the
approach to future null infinity and show how to insure that the
outgoing radiation propagates along flat space null cones which
match up asymptotically with the null cones of the true, physical
space-time.  At every iteration an $O(G^{n})$ gauge
transformation, $\partial\lambda_n^{ab}$, and a small
contribution, $\gamma_{n+1}^{ab} = O(G^{n+1})$, insure that at
large $r$, $h_n^{ab}$ can be written as an expansion in inverse
powers of $r$, times functions of retarded time, $s$, and angle,
${\bf n}$,---in particular there are to be no $\ln r$ terms in
this expansion.  We refer to such an expansion as a proper
expansion in inverse powers of $r$.

We continue to use outgoing-null spherical coordinates,
$(s,r,\theta,\phi)$, but now they are tied to a non-accelerating
world line near the center of the binary system.

At large $r$, $h_1^{ab}$ admits a general multipolar
decomposition just like that presented in
Eqs.\ ({\ref{h1soln}}) and ({\ref{defM}}) and satisfying
Eqs.\ ({\ref{adot}})--({\ref{enodot}}) with the $O(G^2)$ terms
also being $O(r^{-2})$. Thus,
\begin{equation}
  h_1^{ab} \equiv r^{-1} \chi_1^{ab}(s,{\bf n}) + O(r^{-2})
{\label{asymph1}}
\end{equation}
defines $\chi_1^{ab}$, the dominant part of $h_1^{ab}$ at large
$r$.  But at second order, a difficulty immediately arises in
evaluating $p_2^{ab}$.  Namely, $E_1^{ab} = O(r^{-2})$ and Eq.\
({\ref{pnab}}) gives a $r^{-1} \ln r$ term to $p_2^{ab}$ (Theorem
7.2 of Blanchet and Damour{\cite{BlanchetDphil}}).  Such
logarithmic behavior is the signature of a mismatch between the
null cones of the background Minkowskii space and of space-time.
Blanchet{\cite{Blanchet87}} shows that when the $O(r^{-2})$ part
of $E_n^{ab}$ is of a particular form, Eq.\ ({\ref{Enm1=AE}})
below, then the logarithmic terms and the mismatch can be removed
by a gauge transformation.  Our analysis follows
Blanchet{\cite{Blanchet87}} closely, except that we differ on a
choice of gauge for $h_1^{ab}$ and that his analysis involves a
clean separation of the powers of $G$, while our $O(G^n)$ terms
contain further functional dependence on $G$.

Analysis of the definition of $\tau^{ab}(h)$ reveals that if
$k_a \chi_1^{ab}$ were zero then the
offending $r^{-2}$ part of $E_1^{ab}$ would be easy to evaluate.
Generally, $k_a \chi_1^{ab}$ is not zero. In fact, it is
straightforward, but not simple, to see that the
restrictions ({\ref{adot}})--({\ref{enodot}}) imply that
\begin{equation}
  k_a \chi^{ab}_1 =  - v^b ({\cal A}_1 + \dot{\cal D}_1)
         - ({\cal B}_1^b + \dot{\cal H}_1^b),
\end{equation}
where ${\cal A}_1$, ${\cal B}_1^b$, ${\cal D}_1$ and ${\cal H}_1^b$ now refer
to the multipole moments of $h_1^{ab}$ as measured with respect
to the non-accelerating center of the Minkowskii background
geometry.  With foresight, a Lorentz transformation removes the
three-momentum, $({\cal B}_1^b + \dot{\cal H}_1^b)/4$; a gauge
transformation with $\nabla^2 \lambda^a = 0$ (this preserves the
general form of $h_1^{ab}$ and is discussed by
Thorne{\cite{ThorneRMP}}) sets ${\cal D}_1 = 0$; and a second,
preemptive, gauge transformation with $\lambda^a =
(\frac{1}{2}{\cal A}_1\,\ln r,-\frac{1}{4}{\cal A}_1 n^a)$ finally
results in an $h_1^{ab}$ whose lowest non-radiative multipoles
take the simple form at large~$r$
\begin{equation}
  h_1^{ab} \approx \case{1}{2}r^{-1}{\cal A}_1k^a k^b.
\end{equation}
(Interestingly, this is the exact, at all orders, solution for
$h^{ab}$ for a Schwarzschild black hole in outgoing
Eddington-Finkelstein coordinates{\cite{MTW}}.)
With these gauge choices, $h_1^{ab}$ is now in a form such
that
\begin{equation}
  k_a \chi_1^{ab} = 0.
\end{equation}

With this last result, the asymptotic behavior of $E_1^{ab}$ is
dominated by $\tau^{ab}(h_1)$ and is of the form
\begin{equation}
  E_1^{ab} = - r^{-2} k^a k^b \Psi_2(s,{\bf n}) + \bar E_1^{ab}
\end{equation}
where $\bar E_1^{ab} = O(G^2r^{-3})$ and
\begin{equation}
  \Psi_2(s,{\bf n}) = \case{1}{2} \dot \chi_1^{ab} \dot \chi_{1ab}
        - \case{1}{4} \dot \chi^a_{1a} \dot \chi^b_{1b} = O(G^2).
\end{equation}
We note that $\Psi_2$ may be interpreted as being proportional to
the effective energy density of the outgoing gravitational waves.

Generally, this behavior for $E_n^{ab}$ occurs at every order.
Rather than continuing a discussion with a focus on the second
iterative order equations, we switch to the consideration of the
$n$th iterative order and continue following
Blanchet's{\cite{Blanchet87}} analysis closely.

We iteratively assume that $h_{n-1}^{ab}$ has a proper expansion
in inverse powers of $r$ times functions of $s$ and ${\bf n}$,
\begin{equation}
  h_{n-1}^{ab} = r^{-1} \chi_{n-1}^{ab}(s,{\bf n}) + O(Gr^{-2}),
{\label{defzn}}
\end{equation}
with $\delta\chi^{ab}_{n-1}
\equiv\chi^{ab}_{n-1}-\chi^{ab}_{n-2} = O(G^{n-1})$,
\begin{equation}
 k_a \chi_{n-1}^{ab} = 0,
{\label{kdotzn}}
\end{equation}
and that
\begin{equation}
  E_{n-1}^{ab} = - r^{-2} k^a k^b \Psi_n(s,{\bf n}) + \bar
E_{n-1}^{ab}
{\label{Enm1=AE}}
\end{equation}
where $\bar E_{n-1}^{ab} = O(G^n r^{-3})$ and $\Psi_n = O(G^n)$.
First we seek $h_n^{ab} = h_{n-1}^{ab} + \delta h_n^{ab}$
such that $E^{ab}(h_{n-1}+\delta h_n) = O(G^{n+1} r^{-3})$
where this $\delta h_n^{ab}$ differs from that of
\S{\ref{iter}} by a gauge transformation and a small addition and,
also, contains no $\ln r$ term at any
order.  Then we must reconsider the analysis of Eq.\
({\ref{Eabhnplus}}) to account for the difficulties caused by the
$O(G^n r^{-2})$ behavior of $E_{n-1}^{ab}$.

The initial task is to guarantee that
$E_{\text{linear}}^{ab}(\delta h_n) =
                  - E_{n-1}^{ab}+O(G^{n+1}r^{-3})$
so that after the $n$th step
$E_n^{ab}$ will satisfy the incremented version of Eq.\
({\ref{Enm1=AE}}).  The substitution of the Bianchi identity
({\ref{bianchi}}) into Eq.\ ({\ref{divpn}}) results in
\begin{eqnarray}
  \nabla_a p_n^{ab} && = \frac{1}{4\pi} \int \nabla_a'
   [E_{n-1}^{ab} G(x-x') ] \,d^4x'
\nonumber\\ &&
   {} - \frac{1}{4\pi} \int E_{n-1}^{ac} \,
           (\eta_c{}^b \Gamma^d_{da} - \Gamma^b_{ac})
              \,G(x-x')  \,d^4x'.
{\label{divpn1}}
\end{eqnarray}
The second integrand is $O(G^{n+1}r^{-3})$, thus the integral is
$O(G^{n+1})$ and has
a proper expansion in inverse powers of $r$ the leading term of
which matches an outgoing solution of the homogeneous vector wave
equation.  In \S{\ref{iter}}, $q_n^{ab}$ was chosen to cancel just
the $O(G^n)$ part of $\nabla_a p_n^{ab}$; now, $q_n^{ab}$ in Eq.\
({\ref{divqn}}) can be chosen at no additional expense to cancel
this $O(G^{n+1}r^{-1})$ leading term from the second integral as
well.  This results in
\begin{equation}
   \nabla_a (p_n^{ab} + q_n^{ab})
       = r^{-2}\zeta_{n+1}^b(s,{\bf n}) + O(G^{n+1}r^{-3})
{\label{defHb}}
\end{equation}
for some vector $\zeta_{n+1}^b(s,{\bf n}) = O(G^{n+1})$.
Further, if we choose $\gamma_{n+1}^{ab}= O(G^{n+1}r^{-2})$ such
that
\begin{equation}
   k_a \dot \gamma_{n+1}^{ab} = r^{-2} \zeta_{n+1}^b + O(r^{-3}),
{\label{defgamma}}
\end{equation}
Then $p_n^{ab} + q_n^{ab} + \gamma_{n+1}^{ab}$ is nearly what we
seek for $\delta h_n^{ab}$.
With the definition
\begin{equation}
  H_n^{ab} \equiv p_n^{ab} + q_n^{ab} + \gamma_{n+1}^{ab},
\end{equation}
it follows that
\begin{equation}
  \nabla_a H_n^{ab} = O(G^{n+1}r^{-3}).
{\label{divHn}}
\end{equation}
Also
\begin{eqnarray}
  \nabla^2 \gamma_{n+1}^{ab}
  &=& k_c k^c \ddot \gamma_{n+1}^{ab} + O(G^{n+1}r^{-3})
\nonumber\\
    &=&  O(G^{n+1}r^{-3}),
\end{eqnarray}
so that
\begin{equation}
  \nabla^2 H_n^{ab} = E_{n-1}^{ab} + \nabla^2 \gamma_{n+1}^{ab}
       = E_{n-1}^{ab} + O(G^{n+1} r^{-3}).
{\label{del2H}}
\end{equation}
Both Eqs.\ ({\ref{divHn}}) and ({\ref{del2H}}) are used below. And
$E_{\text{linear}}^{ab}(H_n) = -E_{n-1}^{ab} + O(G^{n+1}r^{-3})$
as required, so $H_n^{ab}$ would be the choice for
$\delta h_n^{ab}$ except for the $\ln r$ behavior of $p_n^{ab}$.

An $O(G^n)$ gauge transformation resolves this difficulty and
leaves $E_{\text{linear}}^{ab}$ unchanged.
Blanchet{\cite{Blanchet87}} shows with his Lemma~2.1 that if
\begin{equation}
  \lambda_n^a =  \frac{1}{4\pi}\int
       \frac{1}{2r'^2} k'^a \,
          \int_{-\infty}^{s'} \Psi_n(t,{\bf n'}) dt
  \; G(x-x') \, d^4x'
{\label{deflamn}}
\end{equation}
then
\begin{eqnarray}
  \partial \lambda_n^{ab}= \frac{-1}{4\pi} \int \frac{k'^a
  k'^b}{r'^2} && \Psi_n(s',{\bf n'}) \, G(x-x') \, d^4x'
\nonumber\\ &&
   {} + O(G^n r^{-1})
{\label{gaugetransf}}
\end{eqnarray}
and
\begin{equation}
  \nabla_a \partial \lambda_n^{ab}
            = -\frac{1}{2r^2} k^a
        \int_{-\infty}^{s} \Psi_n(s',{\bf n})\, ds'.
{\label{divlambda}}
\end{equation}
Thus, from Eqs.\ ({\ref{pnab}}), ({\ref{Enm1=AE}}) and
({\ref{gaugetransf}}) the combination $p_n^{ab} + \partial
\lambda_n^{ab}$ is a proper expansion in inverse powers of
$r$.  And, with
\begin{equation}
  \delta h_n^{ab} \equiv p_n^{ab} + q_n^{ab} +
        \partial\lambda_n^{ab} + \gamma_{n+1}^{ab}
\end{equation}
and $h_n^{ab} = h_{n-1}^{ab} + \delta h_n^{ab}$, it follows that
$\delta h_n^{ab}$ has a proper expansion in inverse powers of
$r$, and Eq.\ ({\ref{defzn}}) holds with $n-1 \rightarrow n$.

The iterated versions of Eq.\ ({\ref{kdotzn}}) and
({\ref{Enm1=AE}}) remain to be checked.  From Eqs.\ ({\ref{divHn}})
and ({\ref{divlambda}}) it follows that
$\nabla_a \delta h_n^{ab} = O(G^n r^{-2})$.
But with $\delta h_n^{ab} \equiv r^{-1}
\delta \chi_n^{ab}(s,{\bf n}) + O(G^n r^{-2})$, it must
also be that
\begin{equation}
  \nabla_a \delta h_n^{ab} = - r^{-1} k_a \delta \dot\chi_n^{ab}
          + O(G^n r^{-2}).
\end{equation}
Hence $k_a \delta \dot \chi_n^{ab} = 0$; and, for $n>1$,
$\delta\chi_n^{ab}$ is zero on the initial hypersurface, where
$h_1^{ab}$ matches smoothly onto the initial data, so $k_a
\delta\chi_n^{ab} = 0$ always, and Eq.\ ({\ref{kdotzn}}) holds
with $n-1 \rightarrow n$.

Finally, the analysis of Eq.\ ({\ref{Eabhnplus}}) is
modified by the presence of $\partial\lambda_n^{ab}$ and
$\gamma_{n+1}^{ab}$; but,
\widetext
\begin{eqnarray}
    E^{ab}(h_n) & = &
        [ E^{ab}(h_n) - E^{ab}(h_{n-1}) ]
               + E^{ab}(h_{n-1})
\nonumber\\ & = &
       - \nabla^2 \delta h_n^{cb}
       + \nabla^a\nabla_c \delta h_n^{cb}
       + \nabla^b\nabla_c \delta h_n^{ca}
      -\eta^{ab} \nabla_c \nabla_d \delta h_n^{cd}
     - 16\pi [ \tau^{ab}(h_n) - \tau^{ab}(h_{n-1}) ]
     + E^{ab}(h_{n-1})
\nonumber\\ & = &
       \nabla^a\nabla_c H_n^{cb}
     + \nabla^b\nabla_c H_n^{ca}
     - \eta^{ab} \nabla_c \nabla_d H_n^{cd}
     - 16\pi [ \tau^{ab}(h_n) - \tau^{ab}(h_{n-1}) ]
       + O(G^{n+1}r^{-3})
{\label{Eabhnplus1}}
\end{eqnarray}
%\narrowtext
\noindent where the second equality follows from
the definition of $E^{ab}(h)$,
Eq.\ ({\ref{defE}}).  And the third equality is a consequence
both of $O(G^n)$ gauge invariance of $E_{\text{linear}}^{ab}$
and also of Eq.\ ({\ref{del2H}}).
Now, Eq.\ ({\ref{divHn}}) and the application of
the iterated versions of Eqs.\ ({\ref{defzn}}) and
({\ref{kdotzn}}) to the definition of
$\tau^{ab}(h_n)$ imply that $E^{ab}(h_n)$ is of the form of
Eq.\ ({\ref{Enm1=AE}}) with
\begin{eqnarray}
  \Psi_{n+1}(s,{\bf n})  =  \case{1}{2}
     ( &\dot \chi_n^{ab}& \dot \chi_{nab}
         -  \dot \chi_{n-1}^{ab} \dot \chi_{n-1\,ab})
\nonumber\\ &&
  {} - \case{1}{4} (\dot \chi^a_{na} \dot \chi^b_{nb}
         - \dot \chi^a_{n-1\,a} \dot \chi^b_{n-1\,b} ).
\end{eqnarray}
An iterative step is thus formulated in a manner which leaves
$h_n^{ab}$ expressible as a proper expansion in inverse powers of
$r$ with Eqs.\ ({\ref{defzn}})--({\ref{Enm1=AE}}) holding at every iteration.
And the gravitational waves asymptotically expand out along
constant $s$ surfaces, so the outgoing
null cones of the true space-time
metric asymptotically match up with the flat space null cones.

\section{Conclusions}
We have given a prescription for iteratively improving an
approximate solution to the Einstein equations which could be
carried out, by computer, to any order.  The lowest order
approximation is just the familiar linearized approximation of
general relativity.

The description of the iterative process in the text was intended
to follow a logical order to provide motivation for each part, in
turn, of one full iterative step.  However, in practice the
chronological order is slightly different.  Thus, we now
summarize the entire process with a brief chronological
description of the procedure.

For the initial step we choose $h_1^{ab}$ to be the algebraic sum
of two terms like ${\cal A}_1 v^a v^b /r$, one for each of the two
sources.  Let this $h_1^{ab}$ be considered $h_{n}^{ab}$ and let
the world lines be $z_{n-1}^{ab}$.

In order to iterate the equations of motion, first find
$E_{n}^{ab}$ on each boundary and then find ${\cal R}_{n+1}^b$ and
${\cal P}_{n+1}^b$ from Eqs.\ ({\ref{GFE-9}}) and
({\ref{defPQRS}}). Now, Eqs.\ ({\ref{eom1}}) and ({\ref{fermiw2}})
determine the $\Lambda^{a'}{}_b$ which adjusts the world line to
$z_n^a$ and modifies $h_n^{ab}$ to $h_{n\text{ew}}^{ab}$ with
Eq.\ ({\ref{defhnew}}) while preserving $E^{ab}(h_{n\text{ew}}) =
O(G^{n+1})$.

Next the field equations are iterated by using Eq.\ ({\ref{pnab}})
to determine $p_{n+1}^{ab}$ and Eq.\ ({\ref{qnab}}) for
$q_{n+1}^{ab}$, with the $(n+1)$th order moments satisfying both
Eqs.\ ({\ref{Adot}})--({\ref{Enodot}}) and also appropriate
conditions determined by the physics of the sources within the
boundaries.  The determination of $p_{n+1}^{ab}$ involves solving
the inhomogeneous wave equation for all independent components of
the symmetric tensor.  This is the single, computationally
intensive part of every iterative step.

At this point $E^{ab}(h_n + p_{n+1} + q_{n+1}) = O(G^{n+2})$.
But to preserve proper behavior at future null infinity with the
outgoing null cones of flat space-time matching up asymptotically
with those of the true, physical space-time, a gauge
transformation, $\partial \lambda_{n+1}^{ab}$, from Eq.\
({\ref{deflamn}}) is needed.

In preparation for the next iteration $q_{n+1}^{ab}$ should also
cancel the $O(G^{n+2}r^{-1})$ contribution from the second
integral in Eq.\ ({\ref{divpn1}}).  And $h_{n+1}^{ab}$ should be
changed by a small correction, $\gamma_{n+2}^{ab} =
O(G^{n+2}r^{-2})$, which satisfies Eq.\ ({\ref{defgamma}}).  Now,
$h_{n+1}^{ab} = h_n^{ab} + p_{n+1}^{ab} + q_{n+1}^{ab} +
\partial\lambda_{n+1}^{ab} + \gamma_{n+2}^{ab}$ completes one
full step of the iterative procedure.

The freedom of this iterative process from the restriction of the
harmonic gauge may provide an important aid to its
implementation.  For example the first order $h_1^{ab}$ might be
chosen to be the sum of two terms like ${\cal A} k^a k^b/2r$.  Then
individually each term would be the exact Schwarzschild geometry,
if the source were not accelerating.  And $E^{ab}(h_1)$ would
consist only of linear terms dependent upon the acceleration and
cross terms between the two sources.  This choice for $h_1^{ab}$
would already be an accurate approximation to two Schwarzschild
black holes even near the past event horizon of one hole where
$E^{ab}(h_1)\sim M/R$, with $M$ and $R$ being the mass of and
distance to the {\em companion} hole.

One weakness described in \S{\ref{deltah}} stems from the
inability to treat the conditions at the inner boundaries in a
straightforward manner.  In problems whose focus is on the
emission of gravitational radiation from binary systems, this is
a difficulty only when the system is tight enough that tidal
deformations are important.  To include tidal effects of any sort
it is necessary to solve the internal problem, as described for
example by Damour{\cite{Damour87}}, and thus to obtain the
specific solution of Eqs.\ ({\ref{Adot}})--({\ref{Enodot}}) which
matches the physics of the problem.

Work in progress applies similar methods to a Schwarzschild
background geometry.  In this case it appears that perturbation
analysis ought to yield boundary conditions which can be properly
imposed at the event horizon.  This extension will provide a
better method for the analysis of black hole binary systems.

\section*{Acknowledgments}
This research was supported in part by NASA under grant
NAGW--4864
and was initiated while one of us (S.D.) was
visiting the Jet Propulsion Laboratory, and enjoying the support
and hospitality of Frank Estabrook and Hugo Wahlquist.
And we are grateful to Bernard Whiting for numerous discussions.

\appendix
\section{The retarded Green function.}
{\label{greensfunct}}
The retarded Green function,
\begin{equation}
  G(x-x') = 2 \theta(x^0-x'^0) \delta(\Omega),
\end{equation}
where $\Omega$ is the square of the flat-space interval between
two points,
\begin{equation}
  \Omega(x,x') \equiv \eta_{ab} (x^a - x'^a) (x^b - x'^b),
{\label{GFB-6}}
\end{equation}
is a solution of
\begin{equation}
  \nabla^a \nabla_a G(x-x') = - 4\pi \delta^4(x-x').
{\label{defGreen}}
\end{equation}

For a generic problem we wish to find a particular solution of
\begin{equation}
  \nabla^2 h^{ab} = - \rho^{ab}
\end{equation}
where $\rho^{ab}$ is a multipolar skeleton source on $z^a(s)$.
Thus, if
\begin{equation}
  \rho^{ab}(x) = 4\pi \int M^{abL}(s) \nabla_{L} \delta^4(x-z_s)
        \,ds,
\end{equation}
then
\begin{equation}
  h^{ab}(x) = \int G(x-x') \nabla'_{L} \int M^{abL}(s)
        \delta^4(x'-z_s) \, ds \, d^4x',
\end{equation}
with $\nabla_{d}'$, being the derivative operator with respect to
${x'}^d$.

After integrating by parts $l$ times, changing the derivatives to
be with respect to $x^a$ and integrating over $x'^a$, we have
\begin{equation}
  h^{ab}(x) =
      \int M^{abL}(s) \nabla_{L}G(x-z_s) \, ds,
{\label{wavesoln}}
\end{equation}
or after withdrawing $\nabla_L$ from the integral we have
\begin{equation}
  h^{ab}(x) = \nabla_{L} [r^{-1}M^{abL}(s_x)].
\end{equation}
While this result may appear quite familiar, it is important to
remember that it holds for a source which is moving along some
accelerating world line and has time-changing multipole moments
all the while.  The consequent radiation results from both the
acceleration as well as the varying multipole moments.

\section{A useful Identity} {\label{usefulidentity}}
A useful identity{\cite{Kerr59}} is
\begin{eqnarray}
  \int&& f(s)v^a \nabla_a F(x-z_s) \, ds
     = -\int f(s) \frac{d}{ds} F(x-z_s) \, ds
\nonumber\\
 && = - {\left [ f(s) F(x-z_s) \right |_{-\infty}^{\infty}} +
            \int \frac{df(s)}{ds} F(x-z_s) \, ds.
\end{eqnarray}
For our applications this integral is over all proper time, $s$,
and $F(x-z_s)$ involves the retarded Green function and is zero
except where the past null cone from $x^a$ intersects the world
line. With these conditions the contribution from the limits of
integration is always zero, and
\begin{equation}
  \int f(s) v^a \nabla_a F(x-z_s) \, ds
     = \int \frac{df(s)}{ds} F(x-z_s) \, ds.
{\label{identity}}
\end{equation}
This identity is particularly useful in the reduction leading to
Eq.\ ({\ref{divh1}}).

\section{SSTF decomposition of $N^L$}
{\label{blms}}
By repeated subtraction of the trace parts, $N^L$ is expressed as
\begin{equation}
  N^L = \sum_{m=0}^{[l/2]} b_{l,m}
          f^{(a_1 a_2}  \ldots f^{a_{2m-1}a_{2m}}
          n^{<a_{2m+1}} \ldots n^{a_{l}>)},
{\label{blmdef}}
\end{equation}
for some set of coefficients $b_{l,m}$.  Contraction with
$f_{a_{l-1} a_l}$ and use of Eq.\ ({\ref{blmdef}}) with
$l \rightarrow l-2$ yields
\begin{equation}
  b_{l-2,m} = \frac{(2m+2)(2l-2m-1)}{l(l-1)}b_{l,m+1}.
\end{equation}
It is clear that $b_{l,0} = 1$, and
with elementary methods we find that
\begin{equation}
  b_{l,m} = \frac{l! (2l-4m+1)!!}{2^m m! (l-2m)! (2l-2m+1)!!}.
{\label{blm}}
\end{equation}
Some special values which are of use for determining ${\cal R}^b$ and
${\cal P}^b$ are
\begin{equation}
  b_{2m,m} = 1/(2m+1)
{\label{b2mm}}
\end{equation}
and
\begin{equation}
  b_{1+2m,m} = 3/(2m+3).
{\label{b12mm}}
\end{equation}

\section{Retarded Poincare transformations}
{\label{rpt}}
The retarded Poincare transformation is a little known method for
relating outgoing-null coordinates associated with different
world lines.  This transformation is a mapping from one flat
space-time to another, which transforms one world line, $z^a(s)$,
into a second while preserving the values of the scalar fields
$s$ and $r$; also, the future null cone of each event on the
first world line is mapped to the future null cone of the
corresponding event on the second world line.  Strictly speaking
this transformation is a diffeomorphism from the causal future of
$z^a(s)$ (that is from the set of all events which can be reached
from $z^a(s)$ by a future directed, non-spacelike curve) onto the
causal future of the second world line.  This technicality is
required to allow for the possibility that one world line has
constant acceleration in the distant past, and its causal future
is, thus, not the entire Minkowskii spacetime.

We first describe the mathematical formalism of the retarded
Poincare transformation and then give an application which is
closely related to the analysis of \S{\ref{adjustingh}}.

\subsection{Mathematical formalism}
We start with a given world line $z^a(s)$ in Minkowskii space
covered with the usual Minkowskii coordinates, $x^a$, and
define a coordinate transformation by
\begin{equation}
   y^{a'}  = {\Lambda^{a'}}_b(s_x) x^b + \xi^{a'}(s_x),
{\label{ydef}}
\end{equation}
where ${\Lambda^{a'}}_b$ and $\xi^{a'}$ are functions of $s_x$,
explicitly, and of $x^a$, implicitly, and $\xi^{a'}(s)$ satisfies
Eq.\ ({\ref{xidot}}), below.  The matrix ${\Lambda^{a'}}_b$ is a
time dependent Lorentz transformation, {\it i.e.} it is a matrix
of the general form of a Lorentz boost and a rotation, as
described by Misner {\it et al.}{\cite{MTW}}, but with the boost
and rotation parameters being functions of $s$; also,
${\Lambda^{b}}_{c'}$ is the matrix inverse of
${\Lambda^{a'}}_{b}$.  Thus transformation ({\ref{ydef}}) reduces
to a Lorentz transformation if ${\Lambda^{a'}}_b$ is constant and
$\xi^{a'}=0$; thus, this transformation is a time dependent
generalization of the Lorentz transformation which is reasonably
well behaved in a global sense.

In this section a prime on a base letter identifies a geometrical
object which is most naturally discussed in the $y^{a'}$
coordinate system; a prime on an index refers to the components
of a geometrical object in the $y^{a'}$ coordinate system.

We also define
\begin{equation}
  \eta'^{a'b'}
      \equiv {\Lambda^{a'}}_c {\Lambda^{b'}}_d \eta^{cd}.
\end{equation}
From the algebraic properties of Lorentz transformations we know
that the $y^{a'}$ components of $\eta'^{a'b'}$ are $(-1,1,1,1)$
on the diagonal and zero elsewhere.  While $\eta'^{a'b'}$ is
the usual flat Minkowskii metric of the $y^{a'}$ coordinate
system, it is not the tensor equivalent of $\eta^{ab}$ with
the coordinate transformation ({\ref{ydef}}), because
\begin{equation}
  \eta^{a'b'} =
 \frac{\partial y^{a'}}{\partial x^c}
       \frac{\partial y^{b'}}{\partial x^d} \eta^{cd}
       \neq \eta'^{a'b'}.
\end{equation}
We further define
\begin{equation}
  \eta'_{a'b'} \equiv {\Lambda^c}_{a'} {\Lambda^d}_{b'}
\eta_{cd}.
\end{equation}
Then $\eta'_{a'b'}$ is the  matrix inverse of $\eta'^{b'c'}$
because of the usual algebraic properties of Lorentz
transformations.  And we raise and lower primed indices on
primed tensors with
$\eta'^{b'c'}$ and $\eta'_{b'c'}$; however,  with two different
metrics at hand we rarely raise or lower indices implicitly.
From these definitions it follows that
\begin{equation}
  \eta'_{a'b'} {\Lambda^{b'}}_c  = \eta_{cb}{\Lambda^b}_{a'}
{\label{reletas}}
\end{equation}
along with some index variations of this equation.

For the given world line, $z^a(s)$, we choose $\xi^{a'}(s)$ so
that
\begin{equation}
  \dot \xi^{a'} = - \dot \Lambda^{a'}{}_b z^b(s);
{\label{xidot}}
\end{equation}
this uniquely determines $\xi^{a'}(s)$ up to the addition of a
constant vector. This choice is motivated below, after
Eq.\ ({\ref{dotza}}).

The transformation of the components of tensors is governed by
\begin{eqnarray}
  \frac{\partial y^{a'}}{\partial x^b} & = &
   {\Lambda^{a'}}_b -
     \dot{\Lambda}^{a'}{}_c x^c k_b - \dot{\xi}^{a'} k_b
\nonumber\\ & = &
     \Lambda^{a'}{}_b - r \dot{\Lambda}^{a'}{}_c k^c k_b.
\end{eqnarray}
The inverse transformation is
\begin{equation}
  \frac{\partial x^{b}}{\partial y^{a'}} =
         \Lambda^b{}_{a'}
               - r \dot \Lambda^b{}_{d'} k^{d'} k_{a'};
\end{equation}
the derivation of this inverse involves some of the
results derived below.

We are free to consider the coordinate transformation
({\ref{ydef}}) as a diffeomorphism from one manifold to a second.
Then $\eta^{ab}$ is a flat metric on the $x^a$ manifold, and
$\eta'^{a'b'}$ is a flat metric on the $y^{a'}$ manifold.  With
this point of view, the world line $z^a(s)$ is mapped to a world
line on the $y^{a'}$ manifold by
\begin{equation}
  z'^{a'}(s) = \Lambda^{a'}{}_b z^b(s) + \xi^{a'}(s).
{\label{newline}}
\end{equation}
Then
\begin{equation}
 \dot z'^{a'}(s) = \Lambda^{a'}{}_b \dot z^b(s)
        + \dot \Lambda^{a'}{}_b z^b(s)
        + \dot \xi^{a'};
\label{dotza}
\end{equation}
and with Eq.\ ({\ref{xidot}}), the four-velocities are related by
\begin{equation}
  v'^{a'} \equiv \dot z'^{a'} = {\Lambda^{a'}}_b v^b,
{\label{vnew1}}
\end{equation}
and it follows easily that $\eta'_{a'b'} v^{a'} v^{b'} = -1$
demonstrating that $s$ is the proper time for the world line
$z'^{a'}$ as well.

The future null cone structure of the world line is preserved
under the transformation in the sense that the future null cone
of the event $z^a(s)$ is mapped onto the future null cone of
the event $z'^{a'}(s)$.  For a proof consider the square of the
interval between a generic point on the $y^{a'}$ manifold and a
point on the world line,
\begin{eqnarray}
  &\Omega'&(y^{a'}, z'^{a'})  \equiv
      (y^{a'} - z'^{a'}) (y^{b'} - z'^{b'}) \eta'_{a'b'}
\nonumber\\ & = & \Omega(x^a, z^a)
      + 2 {\Lambda^{a'}}_b (x^b-z^b) \eta'_{a'c'} [\xi^{a'}(s_x)
              - \xi^{a'}(s_z)]
\nonumber\\&&
      + [\xi^{a'}(s_x) - \xi^{a'}(s_z)]
              \eta'_{a'c'} [\xi^{a'}(s_x) - \xi^{a'}(s_z)],
\end{eqnarray}
where the second equality follows from Eqs. ({\ref{ydef}}) and
({\ref{dotza}}).  Thus, if $x^a$ is on the future null cone of
$z^a$ then $s_x = s_z$ and $\Omega(x^a, z^a) = 0$ so that
$\Omega'(y^{a'}, z'^{a'}(s)) = 0$; it follows, then, that
$y^{a'}$ is on the future null cone of $z'^{a'}$ as determined by
the $\eta'_{a'b'}$ metric.

We define a useful scalar field, $r'$, similar to $r$ in
Eq.\ ({\ref{rdef}}), by
\begin{equation}
  r' \equiv -v'_{a'} (y^{a'} - z'^{a'})
 =   - \eta_{cd} v^c [x^d - z^d(s_x)]
 =  r.
\end{equation}
In other words, $r'(y(x)) = r(x)$.

Finally, from
\begin{eqnarray}
  k_a  & = & - \nabla_a s
\nonumber\\ & = &
  ({\Lambda^{b'}}_a - r \dot{\Lambda}^{b'}{}_c k^c k_a)\nabla'_{b'}s,
\nonumber\\ & = &
  ({\Lambda^{b'}}_a - r \dot{\Lambda}^{b'}{}_c k^c k_a) k'_{b'},
{\label{defkprime}}
\end{eqnarray}
and contraction with $k^a$ reveals that $ k^a \Lambda^{b'}{}_a
k'_{b'} = 0$; with some effort it also follows that $\dot
\Lambda^{b'}{}_a k'_{b'} k^a = 0$ and that both
\begin{equation}
  k_a = {\Lambda^{b'}}_a k'_{b'},
\end{equation}
and
\begin{equation}
  k^a = {\Lambda^{a}}_{b'} k'^{b'}.
\end{equation}

With these results at hand, we simplify and summarize the
notation.  The $x^a$ manifold and the $y^{a'}$ manifold have some
similar structures which are distinguished by a prime for the
structure on the $y^{a'}$ manifold. Examples are $z'^{a'}$ and
$\eta'_{a'b'}$ which are similar to $z^{a}$ and $\eta_{ab}$.  But
if we consider the mapping from $x^a$ to $y^{a'}$ to be a
coordinate transformation then we can denote the components of
$\eta'$ in the $x^a$ coordinate system as $\eta'_{ab}$, and as
discussed above $\eta'_{ab} \neq \eta_{ab}$.  But from Eq.\
({\ref{defkprime}}), it does follow that $k_a = k'_{a}$, so that
$k_a$ and $k'_{a'}$ are the same vector in different coordinates;
hence we leave the prime off the base letter $k$; and an index on
$k$ can be raised or lowered by either $\eta$ or $\eta'$.
Additionally, both $s$ and $r$ evaluate to the same scalar fields
on the two different manifolds---and we leave the primes off
these fields as well.

Thus we see that the retarded Poincare transformation described
in Eq.\ ({\ref{ydef}}) is a diffeomorphism which maps one world
line into another while preserving its future null cone and the
values of the scalar fields $s$ and $r$.  Associated with each world line
is a distinct flat metric, $\eta^{ab}$ or $\eta'^{a'b'}$, which
is of the usual Minkowskii diagonal form in the appropriate
(resp. $x^a$ or $y^{a'}$) coordinate system.  We find it most
convenient to be able to move easily between these two manifolds.

It is not difficult to show that the composition of two retarded
Poincare transformations can be described as a single
transformation, and also that any retarded Poincare
transformation has a unique inverse.

For any world line, $z'^{a'}(s)$, there are many retarded
Poincare transformations from $z^{a}(s) = (s,0,0,0)$ to
$z'^{a'}(s)$ which also rotate the coordinate basis
vectors, $e_i^{a'}$.  But, if the $e_i^{a'}$ are Fermi-Walker
transported along $z'^{a'}(s)$, then, as we now
show, the transformation is unique up to an initial rotation of
the basis vectors. Fermi-Walker transport requires that
\begin{equation}
  \dot{e}_i^{a'} = - \Omega'^{a'}{}_{b'} e_i^{b'},
{\label{eidot}}
\end{equation}
where
\begin{equation}
  {\Omega'^{a'}}_{c'} \equiv \dot{v}'^{a'} v'_{c'} -
          v'^{a'} \dot{v}'_{c'}.
{\label{defOmegapr}}
\end{equation}
In addition, for a retarded Poincare transformation the
coordinate basis vectors must also obey
\begin{equation}
  e_i^{a'} = \Lambda^{a'}{}_b e_i^b,
{\label{Lamei}}
\end{equation}
The substitution of Eq.\ ({\ref{Lamei}}) into Eq.\ ({\ref{eidot}}),
along with the orthonormality of the basis vectors, results in
\begin{equation}
  \dot{\Lambda}^{a'}{}_b = - \Omega'^{a'}{}_{c'} \Lambda^{c'}{}_b,
{\label{fermiw}}
\end{equation}
which has a unique solution, given suitable initial conditions.
This last equation, along with Eqs.\ ({\ref{xidot}}) and ({\ref{newline}}),
determines the retarded Poincare transformation that maps
$z^{a}(s) = (s,0,0,0)$ into $z'^{a'}(s)$ with Fermi-Walker
transport of the basis vectors.

In the more general circumstance that $z^a(s)$ as well as
$z'^{a'}(s)$ are arbitrary world lines, Fermi-Walker
transport requires that
\begin{equation}
  \dot\Lambda^{a'}{}_b = - {\Omega'^{a'}}_{c'} {\Lambda^{c'}}_b +
      \Lambda^{a'}{}_c \Omega^c{}_b,
{\label{fermiw2}}
\end{equation}
where $\Omega^{c}{}_{b}$ is defined as in Eq.\
({\ref{defOmegapr}}) but with the primes removed.

\subsection{An application}
{\label{rptscalar}}

An interesting application of the retarded Poincare
transformation, related to the analysis of \S{\ref{adjustingh}},
involves a scalar field which satisfies the wave equation with a
$2^l$-pole source moving along some given world line and with no
incoming radiation at infinity.

For a time dependent source at rest in the $x$ manifold, a simple
expression for the scalar field is
\begin{equation}
  \psi(x) = \sum_{k=0}^l \frac{(-1)^l (l+k)!}{2^k k! (l-k)!}\;
       \frac{{}^{(l-k)}{\cal M}^L(s) \hat{N}_L}{r^{k+1}},
{\label{psiThorne}}
\end{equation}
where ${\cal M}^L(s)$ are the retarded-time dependent SSTF $2^l$-pole
moments of the source, and the prefix superscript of ${\cal M}^L$
denotes differentiation with respect to $s$; this equation is
given by Thorne{\cite{ThorneRMP}} in Eq.\ (2.53a).

More generally, let a retarded Poincare transformation generate a
different world line, $z'^{a'}(s)$, for the source on the
$y^{a'}$ manifold via Eqs.\ ({\ref{ydef}}) and ({\ref{newline}});
then the solution of the scalar wave equation, retaining the same
$2^l$-pole moments as measured by a nearby comoving observer,
is easily written as
\begin{equation}
  \psi'(y) = \nabla'_{L'} ({\cal M}^{L'}(s)/r),
{\label{psi}}
\end{equation}
where $r$ is defined as in Eq.\ ({\ref{rdef}}), and
${\cal M}^{L'} \equiv \Lambda^{L'}{}_{K} {\cal M}^{K}$.  But for an
accelerating world line the evaluation of this right hand side
for $\psi'$ is much more complicated than the right hand side of
Eq.\ ({\ref{psiThorne}}): it involves some terms with up to $l$
derivatives of the velocity and others containing $\dot{v}^l$;
and, even though the source is a $2^l$-pole, $\psi'$ has moments
for the radiation all the way from a monopole up to a
$2^{2l}$-pole.  Thus, although Eq.\ ({\ref{psi}}) looks simple, it is
actually quite difficult to evaluate for an accelerating world line.

Now, consider a new scalar field on the $y^{a'}$ manifold defined
in terms of the scalar field of Eq.\ ({\ref{psiThorne}}) on the
$x$ manifold by
\begin{equation}
    \psi'_{\text{new}}(y) \equiv \psi(x);
\end{equation}
this is not a solution of the scalar wave equation on the
$y^{a'}$ manifold, but for small accelerations it is nearly one.
In fact, inductive evaluation of Eq.\ ({\ref{psi}}) along with
Eqs.({\ref{GFB-4}}) and ({\ref{gradk}}) and the
invariance of $s$ and $r$ under the retarded Poincare
transformation, shows that
\begin{equation}
  \psi'_{\text{new}}(y) = \psi'(y) (1+O(\dot\Lambda)),
\end{equation}
and this is uniformly valid even for large $r$.

A more general analysis reveals that in a similar manner a
retarded Poincare transformation can take a scalar wave solution
for a multipole source moving along one accelerating world line
and generate an approximate solution for the same multipole
source moving along a different world line, as long as the
difference in the accelerations of the world lines is small.
Note that this final limitation restricts neither the total
acceleration of the trajectory nor the size of the boost allowed
in changing the world line, as long as the proper time derivative
of the boost is small.  In this same manner Eq.\ ({\ref{hAnew}})
gives a new self field, $h_{A\text{new}}^{ab}$, when the world
line of the source is changed via a retarded Poincare
transformation.

\bibliographystyle{prsty}

\end{document}